\documentclass[traditabstract]{aa} 
\usepackage{epsfig,amsmath}
\usepackage[dvips]{color}
\usepackage{verbatim}
\usepackage{graphicx}
\usepackage{txfonts}
\usepackage{natbib}
\bibpunct{(}{)}{;}{a}{}{,}
\usepackage{hyperref}
\usepackage{rotating}
\usepackage{multicols,multirow}

\definecolor{Black}{named}{Black}
\definecolor{Red}{named}{Red}
\definecolor{Green}{named}{Green}
\definecolor{Blue}{named}{Blue}

\begin{document}

\title{The ESO UVES Advanced Data Products Quasar Sample - I. Dataset and New $N_{{H}\,{\sc I}}$ Measurements of Damped Absorbers}
\author{Tayyaba Zafar\inst{1}
\and Attila Popping\inst{2}
\and C\'{e}line P\'{e}roux\inst{1}
}

\institute{Aix Marseille Universit\'e, CNRS, LAM (Laboratoire d'Astrophysique de Marseille) UMR 7326, 13388, Marseille, France.
\and International Centre for Radio Astronomy Research (ICRAR), The University of Western Australia, 35 Stirling Hwy, Crawley, WA 6009, Australia.}

\titlerunning{Dataset and New $N_{{H}\,{\sc I}}$ Measurements of Damped Absorbers}
\authorrunning{T. Zafar et al.}

\offprints{tayyaba.zafar@oamp.fr}

\date{Received  / Accepted }

\abstract
{We present here a dataset of quasars observed with the Ultraviolet Visual Echelle Spectrograph (UVES) on the Very Large Telescope and available in the European Southern Observatory UVES Advanced Data Products archive. The sample is made up of a total of 250 high resolution quasar spectra with emission redshifts ranging from  $0.191\leq z_{\rm em}\leq6.311$. The total UVES exposure time of this dataset is 1560 hours. Thanks to the high resolution of UVES spectra, it is possible to unambiguously measure the column density of absorbers with damping wings, down to $N_{{\rm H}\,{\sc \rm I}}\gtrsim10^{19}$ cm$^{-2}$, which constitutes the sub-damped Ly$\alpha$ absorber (sub-DLA) threshold. Within the wavelength coverage of our UVES data, we find 150 damped Ly$\alpha$ systems (DLAs)/sub-DLAs in the range $1.5<z_{\rm abs}<4.7$. Of these 150, 93 are DLAs and 57 are sub-DLAs. An extensive search in the literature indicates that 6 of these DLAs and 13 of these sub-DLAs have their $N_{{\rm H}\,{\sc \rm I}}$ measured for the first time. Among them, 10 are new identifications as DLAs/sub-DLAs. For each of these systems, we obtain an accurate measurement of the {H}\,{\sc i} column density and the absorber's redshift in the range $1.7<z_{\rm abs}<4.2$ by implementing a Voigt profile-fitting algorithm. These absorbers are further confirmed thanks to the detection of associated metal lines and/or lines from members of the Lyman series. In our data, a few quasars' lines-of-sight are rich. An interesting example is towards QSO J0133+0400 ($z_{\rm em} = 4.154$) with six DLAs and sub-DLAs reported.}

\maketitle{}

\section{Introduction}
Among the absorption systems observed in the spectra of quasars, those with the highest neutral hydrogen column density are thought to be connected with the gas reservoir responsible for forming galaxies at high redshift and have deserved wide attention (see review by \citealt{wolfe05}). These systems are usually classified according to their neutral hydrogen column density as 
 damped Ly$\alpha$ systems (hereafter DLAs) with $N_{{\rm H}\,{\sc \rm I}}\ge2\times10^{20}$ atoms cm$^{-2}$ \citep[e.g.,][]{storrie00,wolfe05} and sub-damped Ly$\alpha$ systems (sub-DLAs) with $10^{19}\le N_{{\rm H}\,{\sc \rm I}}\le2\times10^{20}$ atoms cm$^{-2}$ \citep[e.g.,][]{peroux03b}. 

The study of these systems has made significant progress in recent years, thanks to the availability of large sets of quasar spectra with the two-degree field survey (2dF, \citealt{croom01}) and the Sloan Digital Sky Survey (SDSS; \citealt{prochaska05}; \citealt{noterdaeme09}; \citealt{noterdaeme12b}). They have been shown to contain most of the neutral gas mass in the Universe \citep{lanzetta91,lanzetta95,wolfe95} and are currently used to measure the redshift evolution of the total amount of neutral gas mass density \citep{lanzetta91,wolfe95,storrie00,peroux03,prochaska05,noterdaeme09,noterdaeme12b}. In addition, the sub-DLAs may contribute significantly to the cosmic metal budget, which is still highly incomplete. Indeed, only $\sim$20\% of the metals are observed when one adds the contribution of the Ly$\alpha$ forest, DLAs, and galaxies such as Lyman break galaxies \citep[e.g,][]{pettini99,pagel02,wolfe03,pettini04,pettini06,bouche05,bouche06,bouche07}. 

Therefore, to obtain a complete picture of the redshift evolution of both the cosmological neutral gas mass density and  the metal content of the Universe, the less-studied sub-DLAs should be taken into account \citep{peroux03b}. However, these systems cannot be readily studied at low resolution, and only limited samples of high-resolution quasar spectra have been available until now \citep[e.g.,][]{peroux03b,dessauges03,ledoux03,kulkarni07,meiring08,meiring09}.The excellent resolution and large wavelength coverage of UVES allows this less studied class of absorber to be explored.

We have therefore examined the high-resolution quasar spectra taken between February 2000 and March 2007 and available in the UVES \citep{dekker00} Advanced Data Products archive, ending up with a sample of 250 quasar spectra. In this paper we present both the dataset of quasars observed with UVES and the damped absorbers (DLAs and sub-DLAs) covered by these spectra. In addition, we measured column densities of DLAs/sub-DLAs seen in the spectra of these quasars and not reported in the literature. In a companion paper \citep{zafar12b}, we built a carefully selected subset of this dataset to study the statistical properties of DLAs and sub-DLAs, their column density distribution, and the contribution of sub-DLAs to the gas mass density. Further studies, based on specifically designed subsets of the dataset built in this paper, will follow (e.g., studies of metal abundances, molecules).

This work is organized as follows. In \S2, information about the UVES quasar data sample is provided. In \S3, the properties of the damped absorbers are described. This section also summarizes the details of the new column density measurements. In \S4, some global properties of the full sample are presented and lines-of-sight of interest are reported in \S5.  All log values and expressions correspond to log base 10. 

\section{The Quasar Sample}
\subsection{ESO Advanced Data Products}
In 2007, the European Southern Observatory (ESO) managing the 8.2m Very Large Telescope (VLT) observatory has made available to the international community a set of Advanced Data Products for some of its instruments, including the high-resolution UVES\footnote{\url{http://archive.eso.org/eso/eso_archive_adp.html}} instrument. The reduced archival UVES echelle dataset is processed by the ESO UVES pipeline (version 3.2) within the \texttt{MIDAS}  environment with the best available calibration data. This process has been executed by the quality control (QC) group, part of the Data Flow Department. The resulting sample is based on an uniform reprocessing of UVES echelle point source data from the beginning of operations (dated 18$^{\rm th}$ of February 2000) up to the 31$^{\rm st}$ of March 2007. The standard quality assessment, quality control and certification have been integral parts of the process. The following types of UVES data are not included in the product data set: $i)$ data using the image slicers and/or the absorption cell; $ii)$ Echelle data from extended objects and iii) data from the Fibre Large Array Multi Element Spectrograph (FLAMES)/UVES instrument mode.

In general, no distinction has been made between visitor mode (VM) and service mode (SM) data, nor between standard settings and non-standard settings. However, the data reduction was performed only when robust calibration solutions i.e., (``master calibrations") were available. In the UVES Advanced Data Products archive, these calibrations are available only for the standard settings centered on $\lambda$ 346, 390, 437, 520, 564, 580, 600 or 860 nm. For certain ``non-standard" settings, master calibrations were not produced in the first years of UVES operations (until about 2003). These are e.g. 1x2 or 2x3 binnings, or the central wavelengths mentioned above. As a result, the Advanced Data Products database used for the study presented here is not as complete as the ESO UVES raw data archive.

\subsection{Quasars Selection}
The UVES archives do not provide information on the nature of the targets. Indeed, the target names are chosen by the users and only recently does the Phase 2 step propose for the user to classify the targets, but only on a voluntary-basis. Therefore, the first step to construct a sample of quasar spectra out of the Advanced Data Products archive is to identify the nature of the objects. For this purpose we retrieved quasar lists issued from quasar surveys: the Sloan digital sky survey data release 7 (DR7) database\footnote{\url{http://www.sdss.org/dr7/}}, HyperLeda\footnote{\url{http://leda.univ-lyon1.fr/}}, 2dF quasar redshift survey\footnote{\url{http://www.2dfquasar.org/Spec_Cat/2qzsearch2.html}}, Simbad\footnote{\url{http://simbad.u-strasbg.fr/}} and the Hamburg ESO catalogue. The resulting right ascension (RA) and declination (Dec) of the quasars were cross-matched with UVES Advanced Data Products archive within a radius of $15.0''$. The large radius was chosen to overcome possible relative astrometric shifts between the various surveys and the UVES database. Because of this large radius, the raw matched list do not only contain quasars but also other objects such as stars, galaxies, Seyferts. The non-quasar objects have been filtered out by visual inspection of the spectra. The data in an ESO OPC category C (Interstellar Medium, Star Formation and Planetary Systems) and D (Stellar Evolution) are usually targeting galactic objects, but for some cases observers targeted quasars under the same program. The spectra have been visually inspected for those particular cases.

\subsection{Further Data Processing}
In the UVES spectrograph, the light beam from the telescope is split into two arms (UV to Blue and Visual to Red) within the instrument. The spectral resolution of UVES is about $R=\lambda/\Delta\lambda\sim41,400$ when a $1.0''$ slit is used. By varying slit width, the maximum spectral resolution can reach up to $R=80,000$ and $110,000$ for the BLUE and the RED arm, respectively. For each target, individual spectra (most often with overlapping settings) were merged using a dedicated \texttt{Python} code which weights each spectrum by its signal-to-noise ratio. All contributing spectra were regridded to a common frame, with the resolution being that of the spectrum with the highest sampling. When present, the bad pixels were masked to assure that they would not contribute to the merged spectrum. In the regions of overlap the spectra were calibrated to the same level before being error-weighted and merged. Particular attention was given to ``step" features in the quasar continua and a visual search has identified and corrected these features when they corresponded to the position in between two orders of the Echelle spectrum. In the merging process for each individual spectrum, a radial velocity correction for barycentric and heliocentric motion (using heliocentric correction values from the files header) was applied. A vacuum correction on the wavelength was also applied.

The resulting list comprises 250 quasar spectra. The number of individual spectra used to produce the co-added spectrum, Simbad $V$-band magnitudes, together with total exposure time in seconds for each target, are provided in Table \ref{proptab}. Throughout the paper, this sample obtained from the ESO UVES Advanced Data Products facility is called EUADP sample. The total VLT exposure time of this dataset is $T_{\rm tot}=1560$ hours. 
 
 In the case of close pairs of quasars or gravitationally-lensed quasars, we have separated the objects if different slit-positions were used but only analyzed the brightest object if these objects were aligned along the slit. Our sample contains two lensed quasars: QSO B0908+0603 (double system) and QSO B1104-181 (quadruple system). In the former case two objects, and, in the latter case, three objects were aligned on the slit \citep{lopez07}. In these cases we only analyzed the brightest objects. Our sample contains four quasar pairs: QSO J0008-2900 \& QSO J0008-2901 (separated by $1.3'$), J030640.8-301032 \& J030643.7-301107 (separated by $\sim$0.85$'$), QSO B0307-195A \& B (separated by $\sim$1$'$), and QSO J1039-2719 \& QSO B1038-2712 (separated by $17.9'$). For these eight objects, eight different slit-positions were used.
 
      \begin{figure}
   \centering
   {\includegraphics[width=\columnwidth,clip=]{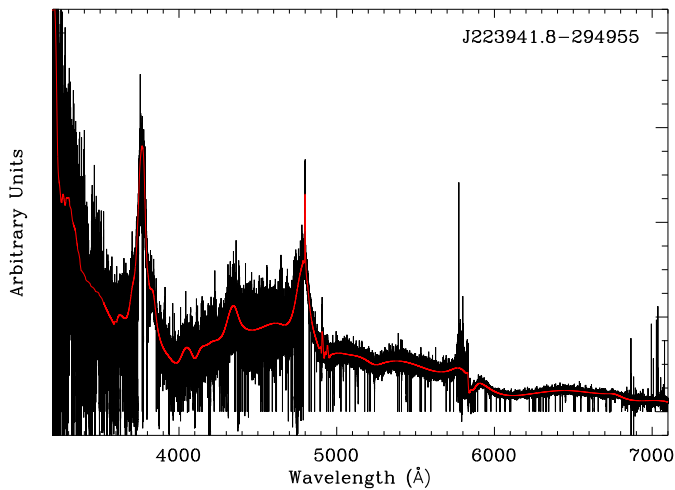}} 
      \caption{An example of a UVES quasar spectrum of J223941.8-294955 at $z_{em}=2.102$. The quasar continuum (red overlay) is fitted using a spline function.}
          \label{conti}
   \end{figure}

\section{The DLA/sub-DLA Sample}
The quasar spectra were normalized to unity within the \texttt{MIDAS} environment. For each quasar the local continuum was determined in the merged spectrum by using a spline function to smoothly connect the regions free of absorption features (see Fig. \ref{conti}). The final normalized spectra used for column density analysis were obtained by dividing merged spectra by these continua. During normalization, the artifact from residual fringes (especially in standard setting centered on $\lambda$ 860nm) and spectral gaps are also included in the spline fitting. In order to secure all DLAs and sub-DLAs in a given spectrum, we proceeded as in \citet{lanzetta91} and used an automated \texttt{Python} detection algorithm. This code builds an equivalent width spectrum over 400 pixel wide boxes ($\sim$12\,\AA\ for 0.03\,$\AA$ pixel$^{-1}$) blueward of the quasar's Ly$\alpha$ emission and flags regions of the spectra where the observed equivalent width exceeds the sub-DLA definition (i.e. $N_{{\rm H}\,{\sc \rm I}}\gtrsim10^{19}$ cm$^{-2}$). This candidate list of DLAs and sub-DLAs is further supplemented by visual inspection done by TZ and CP. The DLAs and sub-DLAs have been confirmed by looking for associated metal lines and/or higher members of the Lyman series. This resulted in 150 DLAs/sub-DLAs with $1.6<z_{\rm abs}<4.7$ for which Ly$\alpha$ is covered in the spectra of these quasars. This method is complete down to the sub-DLA definition of $N_{{\rm H}\,{\sc \rm I}}=10^{19}$ equivalent to $\rm EW_{\rm rest}=2.5$\,$\AA$ based on a curve of growth analysis \citep{dessauges03}.

\subsection{$N_{{\rm H}\,{\sc \rm I}}$ measurements of DLAs and sub-DLAs}
An extensive search in the literature was undertaken to identify which of these damped absorbers were already known. Of the 150 DLAs/sub-DLAs that we have identified above, 131 (87 DLAs and 44 sub-DLAs) have their $N_{{\rm H}\,{\sc \rm I}}$ already reported in the literature. Of the remaining 19 (6 DLAs and 13 sub-DLAs), 10 (3 DLAs and 7 sub-DLAs) are new identifications (see Table \ref{dlamet} and \S\ref{notes} for details on each system).

For damped absorbers, the Lorentzian component of the Voigt profile results in pronounced damping wings, allowing precise determination of $N_{{\rm H}\,{\sc \rm I}}$ down to the sub-DLA definition at high-resolution. The neutral hydrogen column density measurements of these absorbers are determined by fitting a Voigt profile to the Ly$\alpha$ absorption line. The fits were performed using the $\chi^2$ minimization routine \texttt{FITLYMAN} package in \texttt{MIDAS} \citep{fontana}. Laboratory wavelengths and oscillator strengths from \citet{morton2003} were used. The global fit returns the best fit parameters for central wavelength, column density, and Doppler turbulent broadening, as well as errors on each quantity. The central redshift was left as a free parameter except when no satisfactory fit could be found in which case the strongest component of the metal line was chosen as central redshift. The Doppler turbulent parameter $b$-value was usually left as a free parameter and sometimes fixed because of low signal-to-noise or multiple DLAs at 30 km s$^{-1}$ (for DLAs) or $20$ km s$^{-1}$ (for sub-DLAs). The $N_{{\rm H}\,{\sc \rm I}}$ fit is performed using the higher members of the Lyman series, in addition to Ly$\alpha$, where these are available. For fitting $N_{{\rm H}\,{\sc \rm I}}$, we usually used the components from $\ion{O}{i}$. The other low ionization line components are used to fit $N_{{\rm H}\,{\sc \rm I}}$ for the cases where there is no $\ion{O}{i}$ covered by our data. Table \ref{dlamet} summarizes the properties of the DLAs/sub-DLAs for which we obtained ${\rm H}\,{\sc \rm I}$ column density for the first time and provides quasar emission redshifts, absorption redshifts, {H}\,{\sc i} column densities and metal line lists.
 
Moreover, the majority of the $N_{{\rm H}\,{\sc \rm I}}$ measurements of the DLAs/sub-DLAs towards the 250 EUADP quasars comes from high-resolution data mostly from UVES or Keck/HIRES. In 7 cases, we cover the DLA/sub-DLA in our data but the $N_{{\rm H}\,{\sc \rm I}}$ in the literature is obtained from low/moderate resolution spectra. For these 7 cases, we refitted the DLA/sub-DLA using the EUADP data and new $N_{{\rm H}\,{\sc \rm I}}$ are reported in Table \ref{newNH}. For most of the cases, we find consistent results with the low resolution studies. In the case of QSO\,B1114-0822, \citet{storrie00} reported a DLA with log $N_{{\rm H}\,{\sc \rm I}}=20.3$ at $z_{\rm abs}=4.258$ while we find a sub-DLA with log $N_{{\rm H}\,{\sc \rm I}}=20.02\pm0.12$.

 \addtocounter{table}{+1}
\begin{table*}
\caption{List of 6 damped and 13 sub-damped absorbers with new $N_{{\rm H}\,{\sc \rm I}}$ measurements. The details of these absorbers are provided in the columns as (1) Simbad quasar name, (2) emission redshift, (3) absorption redshift, (4) {H}\,{\sc i} column density with corresponding errors, (5) members of Lyman series covered, (6) metal lines associated with the system, and (7) is the absorber a new identification as DLA/sub-DLA?}
\label{dlamet}
\centering
\setlength{\tabcolsep}{3pt}
\begin{tabular}{@{} l c c c c c c c@{}}
\hline
 Quasar &  $z_{\rm em}$ &  $z_{\rm abs}$ &  log $N_{{\rm H}\,{\sc \rm I}}$ &  Ly series &  Metals & New? \\
& & &  cm$^{-2}$ & \\
 \hline
QSO J0008-2900 & 2.645 & 2.254 & $20.22\pm0.10$ & Ly$\beta$ & {O}\,{\sc i}, {Fe}\,{\sc ii}, {Si}\,{\sc ii}, {C}\,{\sc ii}, {Mg}\,{\sc ii}, {Al}\,{\sc iii}, {Si}\,{\sc iv} & no \\
QSO J0008-2901 & 2.607 & 2.491 & $19.94\pm0.11$ & Ly$\gamma$ & {O}\,{\sc i}, {Fe}\,{\sc ii}, {Si}\,{\sc ii}, {C}\,{\sc ii}, {Al}\,{\sc ii}, {Si}\,{\sc iv} & no  \\ 
QSO J0041-4936 & 3.240 & 2.248 & $20.46\pm0.13$ & Ly$\beta$ & {O}\,{\sc i}, {Fe}\,{\sc ii}, {Si}\,{\sc ii}, {C}\,{\sc ii}, {Al}\,{\sc ii}, {Zn}\,{\sc ii}, {Al}\,{\sc iii}, {C}\,{\sc iv} & no \\
QSO B0128-2150 & 1.900 & 1.857 & $20.21\pm0.09$ & Ly$\alpha$ & {O}\,{\sc i}, {Fe}\,{\sc ii}, {Si}\,{\sc ii}, {C}\,{\sc ii}, {Al}\,{\sc iii} & no \\ 
J021741.8-370100 & 2.910 & 2.429 & $20.62\pm0.08$ & Ly$\beta$ & {O}\,{\sc i}, {Si}\,{\sc ii} & no \\
$\cdots$ & $\cdots$ & 2.514 & $20.46\pm0.09$ & Ly$\beta$ & {Fe}\,{\sc ii}, {Si}\,{\sc ii} & no \\
J060008.1-504036	& 3.130  & 2.149 & $20.40\pm0.12$ &  Ly$\alpha$ & {O}\,{\sc i}, {Fe}\,{\sc ii}, {Si}\,{\sc ii}, {C}\,{\sc ii}, {Al}\,{\sc ii}, {Al}\,{\sc iii} & yes \\
QSO B0952-0115 & 4.426  &  3.476 & $20.04\pm0.07$ &  Ly$\alpha$ & {Fe}\,{\sc ii}, {Si}\,{\sc ii}, {Al}\,{\sc ii}, {Al}\,{\sc iii}, {C}\,{\sc iv}  & yes \\
Q1036-272 & 3.090 &  2.792 & $20.65\pm0.13$ & Ly$\beta$ & {O}\,{\sc i}, {Fe}\,{\sc ii}, {Si}\,{\sc ii}, {Al}\,{\sc iii} & yes \\
QSO B1036-2257 & 3.130 & 2.533 & $19.30\pm0.10$ & Ly$\alpha$ & {Fe}\,{\sc ii}, {Si}\,{\sc ii}, {C}\,{\sc ii}, {Al}\,{\sc iii}, {Si}\,{\sc iv}, {C}\,{\sc iv} & no \\
QSO B1036-268 & 2.460 & 2.235 & $19.96\pm0.09$ & Ly$\beta$ & {O}\,{\sc i}, {Fe}\,{\sc ii}, {Si}\,{\sc ii}, {C}\,{\sc ii}, {Al}\,{\sc ii}, {Mg}\,{\sc ii}, {Al}\,{\sc iii}, {Si}\,{\sc iv}, {C}\,{\sc iv} & yes  \\
LBQS 1232+0815 & 2.570 & 1.720 & $19.48\pm0.13$ &  Ly$\alpha$ & {O}\,{\sc i}, {Fe}\,{\sc ii}, {Si}\,{\sc ii}, {Al}\,{\sc iii}, {Si}\,{\sc iv}, {C}\,{\sc iv} & yes \\
QSO J1330-2522 & 3.910 & 2.654 & $19.56\pm0.13$ & Ly$\alpha$ &  {Si}\,{\sc ii}, {Al}\,{\sc ii}, {Al}\,{\sc iii}, {Si}\,{\sc iv}, {C}\,{\sc iv} & yes \\
QSO J1356-1101 & 3.006 & 2.397 & $19.88\pm0.09$ & Ly$\alpha$ & {O}\,{\sc i}, {Fe}\,{\sc ii}, {Si}\,{\sc ii} & yes \\
QSO J1723+2243 & 4.520 & 4.155 & $19.23\pm0.12$ & Ly$\gamma$ & metal lines blended or not covered & yes \\
LBQS 2114-4347 & 2.040 & 1.912 & $19.50\pm0.10$ & Ly$\alpha$ & {O}\,{\sc i}, {Fe}\,{\sc ii}, {Si}\,{\sc ii}, {C}\,{\sc ii}, {Al}\,{\sc ii}, {Mg}\,{\sc ii}, {Si}\,{\sc iv}, {C}\,{\sc iv} & yes \\
J223941.8-294955 & 2.102  & 1.825 & $19.84\pm0.14$ & Ly$\alpha$ & {O}\,{\sc i}, {Fe}\,{\sc ii}, {Si}\,{\sc ii}, {C}\,{\sc ii}, {Al}\,{\sc ii}, {Mg}\,{\sc ii}, {Al}\,{\sc iii}, {Si}\,{\sc iv}, {C}\,{\sc iv} & no \\
QSO B2318-1107 & 2.960 & 1.629 & $20.52\pm0.14$ & Ly$\alpha$ & {Fe}\,{\sc ii}, {Si}\,{\sc ii}, {C}\,{\sc ii}, {Al}\,{\sc ii}, {Al}\,{\sc iii}, {Si}\,{\sc iv} & yes \\
QSO B2342+3417 & 3.010  & 2.940 & $20.18\pm0.10$ & Ly$\gamma$ & limited red wavelength coverage & no \\
\hline
\end{tabular}
\end{table*}

\subsection{Notes on Individual Objects}\label{notes}
In this section, we provide details on the DLAs and sub-DLAs in the EUADP sample for which ${\rm H}\,{\sc \rm I}$ column density is determined in this work. Best fit parameters of the Voigt profile fits to the {H}\,{\sc i} absorption lines are detailed below.

      \begin{figure}
   \centering
   {\includegraphics[width=\columnwidth,height=7.5cm,clip=]{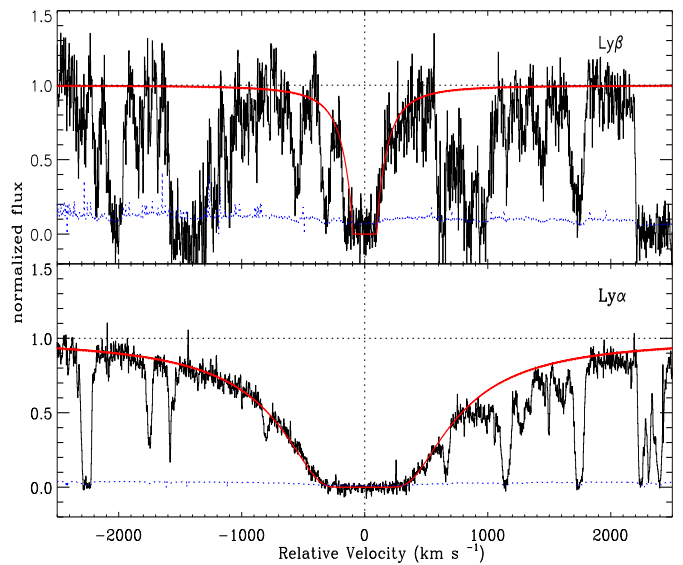}} 
      \caption{The sub-DLA detected towards QSO J0008-2900 at $z_{\rm abs}=2.254$. The solid red line shows the Voigt profile fit to the sub-DLA with an inferred column density of log $N_{{\rm H}\,{\sc \rm I}}=20.22\pm0.10$. The system is detected down to Ly$\beta$. Here and in the following figures, the dotted blue line represents 1$\sigma$ error on the spectrum. The vertical dotted line is the adopted zero velocity corresponding to the redshift of the sub-DLA. The horizontal line is at a level of one.}
          \label{J0008-2900}
   \end{figure}

      \begin{figure}
   \centering
   {\includegraphics[width=\columnwidth,height=9cm,clip=]{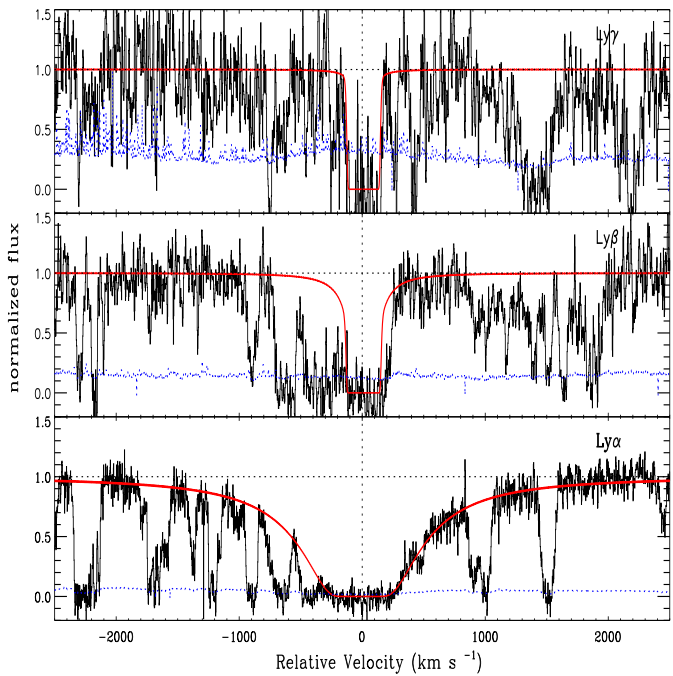}}
      \caption{The sub-DLA detected towards QSO J0008-2901 at $z_{\rm abs}=2.491$. The solid red line corresponds to the Voigt profile fit to the sub-DLA with an inferred column density of log $N_{{\rm H}\,{\sc \rm I}}=19.94\pm0.11$. The system is detected down to Ly$\gamma$.}
          \label{J0008-2901}
   \end{figure}

 \begin{table}
\begin{minipage}[t]{\columnwidth}
\caption{New high-resolution $N_{{\rm H}\,{\sc \rm I}}$ measurements of DLAs/sub-DLAs previously observed at low/medium resolution.}
\label{newNH}
\centering
\setlength{\tabcolsep}{3pt}
\begin{tabular}{l c c c c }
\hline
Quasar & \multicolumn{2}{c}{High resolution} & \multicolumn{2}{c}{Low/medium resolution} \\
& $z_{\rm abs}$ & log $N_{{\rm H}\,{\sc \rm I}}$ &  Survey/ &  log $N_{{\rm H}\,{\sc \rm I}}$  \\
 &  & cm$^{-2}$  & Instrument  & cm$^{-2}$ \\
\hline
J004054.7-091526 & 4.538 & $20.20\pm0.09$ & SDSS & $20.22\pm0.26$$^{1}$ \\
$\cdots$ & 4.740 & $20.39\pm0.11$ & SDSS & $20.55\pm0.15$$^{2}$ \\
QSO B1114-0822 & 4.258 & $20.02\pm0.12$ & WHT & 20.3$^{3}$ \\
J115538.6+053050 & 2.608 & $20.37\pm0.11$ & SDSS & $20.47\pm0.23$$^{1}$ \\
$\cdots$ & 3.327 & $21.00\pm0.10$ & SDSS  & $20.91\pm0.27$$^{1}$ \\
LBQS 2132-4321 & 1.916 & $20.74\pm0.09$ & LBQS & 20.70$^{3}$ \\
J235702.5-004824 & 2.479 & $20.41\pm0.08$ & SDSS & 20.45$^{2}$ \\
\hline
\end{tabular}
\end{minipage}
\vspace{0.1cm}
{\bf References:} (1) \citet{noterdaeme09};
(2) \citet{prochaska08}; 
(3)\citet{storrie00}
\end{table}

\begin{enumerate}
\item QSO J0008-2900 ($z_{\rm em}=2.645$). The quasar was discovered during the course of the 2dF quasar redshift survey \citep{croom01}. An {H}\,{\sc i} absorber at $z=2.253$ is reported by \citet{tytler09}. We find that the absorber is a sub-DLA with log $N_{{\rm H}\,{\sc \rm I}}=20.22\pm0.10$ and $b=28.5\pm2.2$ km s$^{-1}$ at $z_{\rm abs}=2.254$. The Lyman series lines down to Ly$\beta$ are fitted together. In the red part of the spectrum metal lines from  {O}\,{\sc i} $\lambda$ 1302, {Fe}\,{\sc ii} $\lambda\lambda\lambda\lambda$ 2344, 2374, 2382, 2586, {Si}\,{\sc ii} $\lambda\lambda\lambda\lambda$ 1260, 1304, 1526, 1808, {C}\,{\sc ii} $\lambda$ 1334, {Mg}\,{\sc ii} $\lambda\lambda$ 2796, 2803, {Al}\,{\sc iii} $\lambda\lambda$ 1854, 1862, and {Si}\,{\sc iv} $\lambda\lambda$ 1393, 1402 are detected at the redshift of the sub-DLA. Fig. \ref{J0008-2900} shows our best fit result of the {H}\,{\sc i} lines. 

       \begin{figure}
   \centering
         {\includegraphics[width=\columnwidth,height=7.5cm,clip=]{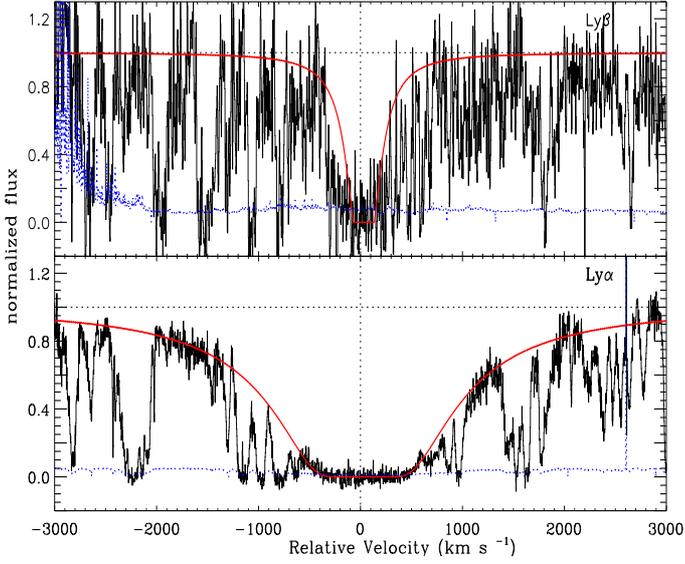}} 
      \caption{The DLA detected towards QSO J0041-4936 at $z_{\rm abs}=2.248$. For plotting purposes the Ly$\beta$ region of the spectrum is smoothed with a boxcar average of 2.0 pixels. The solid red line represents the Voigt profile fit to the DLA with an inferred column density of log $N_{{\rm H}\,{\sc \rm I}}=20.46\pm0.13$. The system is detected down to Ly$\beta$.}
          \label{J0041-4936}
   \end{figure}

      \begin{figure}
   \centering
            {\includegraphics[width=0.9\columnwidth,clip=]{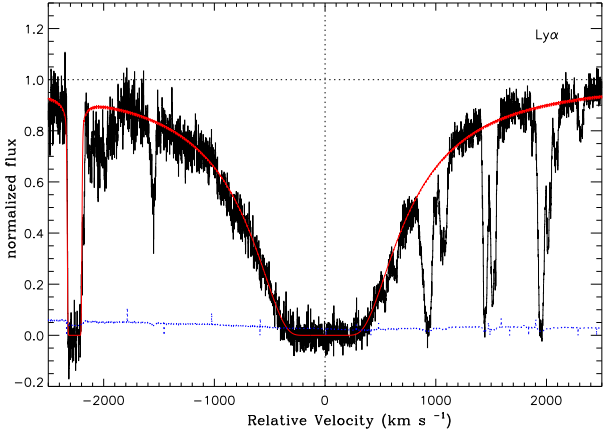}} 
      \caption{The sub-DLA detected towards QSO B0128-2150 at $z_{\rm abs}=1.857$. The solid red line shows the Voigt profile fit to the sub-DLA with an inferred column density of log $N_{{\rm H}\,{\sc \rm I}}=20.21\pm0.09$.}
          \label{B0128-2150}
   \end{figure}

\item QSO J0008-2901 ($z_{\rm em}=2.607$). The quasar was also discovered during the course of the 2dF quasar redshift survey \citep{croom01}. An {H}\,{\sc i} absorber at $z=2.491$ is reported by \citet{tytler09}. We find that the absorber at $z_{\rm abs}=2.491$ is a sub-DLA with log $N_{{\rm H}\,{\sc \rm I}}=19.94\pm0.11$ and $b=39\pm3.5$ km s$^{-1}$ detected down to Ly$\gamma$. Metal lines from {O}\,{\sc i} $\lambda$ 1302, {Fe}\,{\sc ii} $\lambda\lambda\lambda\lambda$ 2344, 2374, 2382, 2586, {Si}\,{\sc ii} $\lambda\lambda\lambda$ 1260, 1304, 1808, {C}\,{\sc ii} $\lambda$ 1334, {Al}\,{\sc ii} $\lambda$ 1670, and {Si}\,{\sc iv} $\lambda\lambda$ 1393, 1402 are detected in the red part of the spectrum. Fig. \ref{J0008-2901} shows our best fit result of the {H}\,{\sc i} lines. 

\item QSO J0041-4936 ($z_{\rm em}=3.240$). A damped absorber is known in this quasar from the Cal\'{a}n Tololo survey \citep{maza95}. From a low-resolution spectrum, \citet{lopez01} measured the equivalent width of the {H}\,{\sc i} absorption line to be ${\rm EW_{obs}}=34.60\,\AA$ but state that they cannot measure the {H}\,{\sc i} column density due to the limited spectral resolution. Using the high-resolution UVES spectrum, we are able to measure the column density of the DLA to be log $N_{{\rm H}\,{\sc \rm I}}=20.46\pm0.13$ and $b=29\pm3.9$ km s$^{-1}$ at $z_{\rm abs}=2.248$ detected down to Ly$\beta$. Metal lines from {O}\,{\sc i} $\lambda$ 1302, {Fe}\,{\sc ii} $\lambda\lambda$ 1608, 1611, {Si}\,{\sc ii} $\lambda\lambda\lambda\lambda$ 1260, 1304, 1526, 1808, {C}\,{\sc ii} $\lambda$ 1334, {Al}\,{\sc ii} $\lambda$ 1670, {Zn}\,{\sc ii} $\lambda$ 2026, {Al}\,{\sc iii} $\lambda\lambda$ 1854, 1862, and {C}\,{\sc iv} $\lambda\lambda$ 1548, 1550 associated with the DLA are detected in the red part of the spectrum. Fig. \ref{J0041-4936} shows the best fit result of the {H}\,{\sc i} lines.

      \begin{figure}
   \centering
   {\includegraphics[width=\columnwidth,height=7.5cm,clip=]{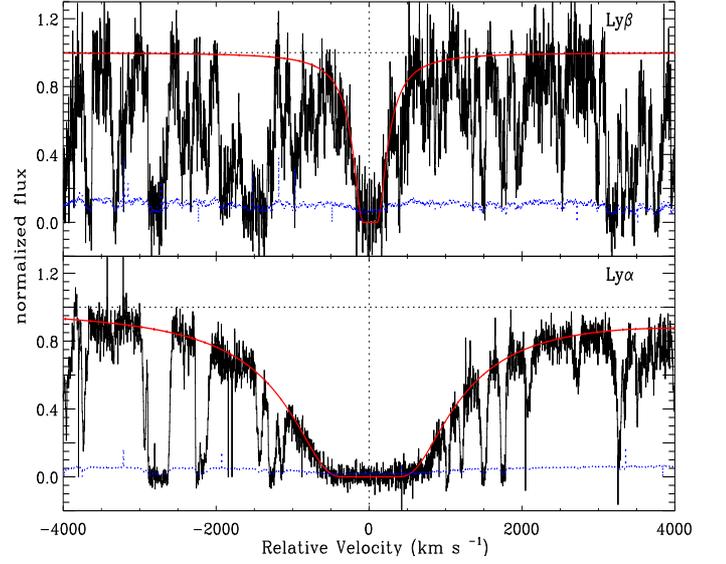}}
      \caption{The DLA detected towards J021741.8-370100 at $z_{\rm abs}=2.429$. The solid red line represents the Voigt profile fit to the DLA with an inferred column density of log $N_{{\rm H}\,{\sc \rm I}}=20.62\pm0.08$. The system is detected down to Ly$\beta$.}
          \label{J021741.8-370100a}
   \end{figure}

\item QSO B0128-2150 ($z_{\rm em}=1.900$). This quasar was discovered during the course of the Montr\'{e}al-Cambridge-Tololo survey \citep{lamontagne00}. Two absorbing systems at $z_{\rm abs}=1.64$ and 1.85 are reported in the UVES observing proposal. We find that the system at $z_{\rm abs}=1.857$ is a sub-DLA with log $N_{{\rm H}\,{\sc \rm I}}=20.21\pm0.09$ and $b=25\pm2.6$ km s$^{-1}$. Metal lines from {O}\,{\sc i} $\lambda$ 1302, {Fe}\,{\sc ii} $\lambda\lambda\lambda$ 2344, 2374, 2382, {Si}\,{\sc ii} $\lambda\lambda\lambda$ 1260, 1304, 1808, {C}\,{\sc ii} $\lambda$ 1334, and {Al}\,{\sc iii} $\lambda\lambda$ 1854, 1862 at the redshift of the absorber are observed in the red part of the spectrum. Fig. \ref{B0128-2150} shows our best fit result of the {H}\,{\sc i} column density. The absorbing system at $z_{abs}=1.64$ is below sub-DLA limit.

    \begin{figure}
   \centering
      {\includegraphics[width=\columnwidth,height=7.5cm,clip=]{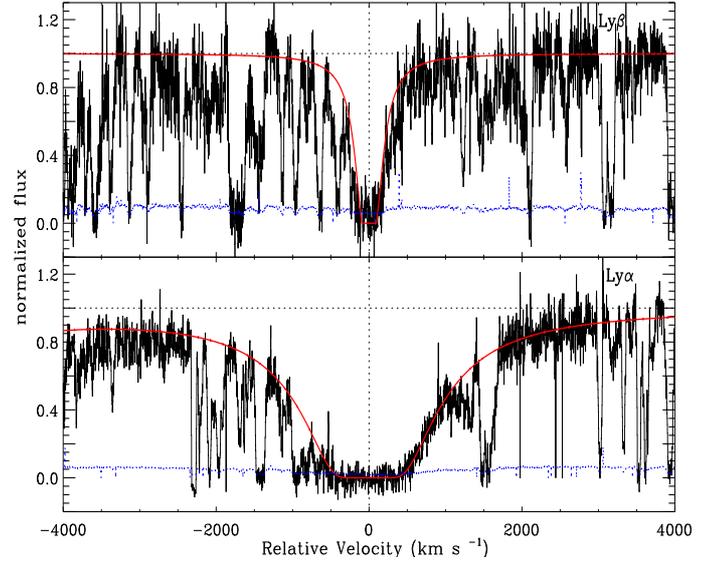}} \\
      \caption{The DLA detected towards J021741.8-370100 at $z_{\rm abs}=2.514$. The solid red line corresponds to the Voigt profile fit to the DLA with an inferred column density of log $N_{{\rm H}\,{\sc \rm I}}=20.46\pm0.13$. The system is detected down to Ly$\beta$.}
         \label{J021741.8-370100b}
   \end{figure}

\item J021741.8-370100 ($z_{\rm em}=2.910$). Damped absorbers have been known in this quasar from the Cal\'{a}n Tololo survey \citep{maza96}. The column densities for these DLAs have not been reported before. We measure log $N_{{\rm H}\,{\sc \rm I}}=20.62\pm0.08$ at $z_{\rm abs}=2.429$ and log $N_{{\rm H}\,{\sc \rm I}}=20.46\pm0.09$ at $z_{\rm abs}=2.514$. Both absorbers are seen down to Ly$\beta$. The $b$ parameter is fixed to $b=30$ km s$^{-1}$ for both absorbers. Due to limited wavelength coverage only a few metal lines associated with the absorbers are seen in the spectrum. Metal lines from {O}\,{\sc i} $\lambda$ 1302 and {Si}\,{\sc ii} $\lambda\lambda\lambda\lambda$ 1190, 1193, 1260, 1304 are covered for the absorber at $z_{\rm abs}=2.429$. The lines from {Fe}\,{\sc ii} $\lambda$ 1144 and {Si}\,{\sc ii} $\lambda\lambda\lambda$ 1190, 1193, 1260 associated with the absorber at $z_{\rm abs}=2.514$ are covered in the spectrum. Fig. \ref{J021741.8-370100a} and Fig. \ref{J021741.8-370100b} show our best fit of {H}\,{\sc i} lines for absorbers at $z_{\rm abs}=2.429$ and $z_{\rm abs}=2.514$ respectively.

    \begin{figure}
   \centering
      {\includegraphics[width=0.9\columnwidth,clip=]{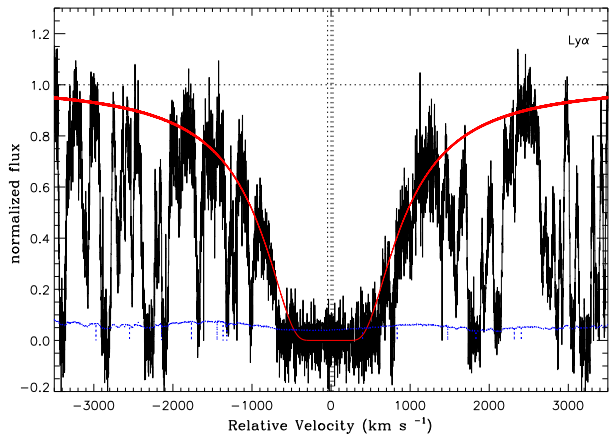}} \\
      \caption{The DLA detected towards J060008.1-504036 at $z_{\rm abs}=2.149$. The solid red line shows the Voigt profile fit to the DLA with a total column density of log $N_{{\rm H}\,{\sc \rm I}}=20.40\pm0.12$ using velocity components at 24, 0, -44 km s$^{-1}$.}
          \label{J060008.1-504036}
   \end{figure}

    \begin{figure}
   \centering
      {\includegraphics[width=0.9\columnwidth,clip=]{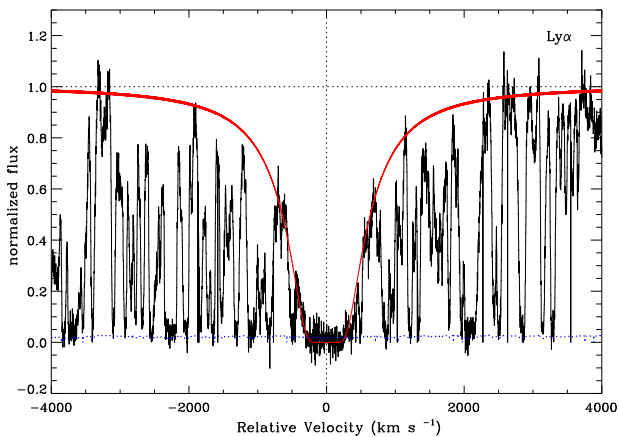}} \\
      \caption{The sub-DLA detected towards J0952-0115 at $z_{\rm abs}=3.476$. The solid red line corresponds to the Voigt profile fit to the sub-DLA with an inferred column density of log $N_{{\rm H}\,{\sc \rm I}}=20.04\pm0.07$.}
          \label{B0952-0115}
   \end{figure}

\item J060008.1-504036 ($z_{\rm em}=3.130$). This quasar was discovered during the course of the Cal\'{a}n Tololo survey \citep{maza96}. No detailed analysis of this quasar has been published. A Lyman limit system at $z_{\rm LLS}=3.080$ is seen in the spectrum. We identified a DLA at $z_{\rm abs}=2.149$. The column density of the DLA is fitted using three strong components (at 24, 0, and -44 km s$^{-1}$) seen in $\ion{O}{i}$, resulting in a total column density of log $N_{{\rm H}\,{\sc \rm I}}=20.40\pm0.12$ with $b=20$ km s$^{-1}$ fixed for each component. Metal absorption lines from {O}\,{\sc i} $\lambda$ 1302, {Fe}\,{\sc ii} $\lambda\lambda$ 1608, 1611, {Si}\,{\sc ii} $\lambda\lambda\lambda\lambda$ 1260, 1304, 1526, 1808, {C}\,{\sc ii} $\lambda$ 1334, {Al}\,{\sc ii} $\lambda$ 1670, and {Al}\,{\sc iii} $\lambda\lambda$ 1854, 1862 at the redshift of the DLA are covered in the red part of the spectrum. Fig. \ref{J060008.1-504036} shows our best fit of the neutral hydrogen column density of the DLA at $z_{\rm abs}=2.149$.

    \begin{figure}
   \centering
      {\includegraphics[width=\columnwidth,height=7.5cm,clip=]{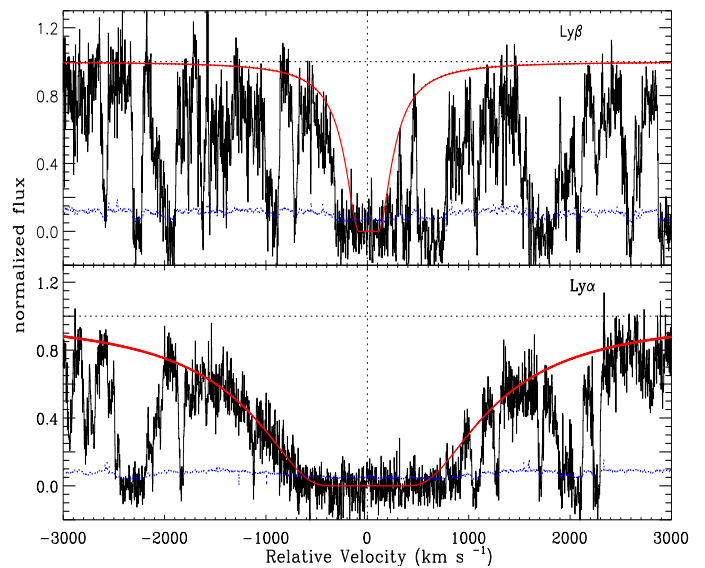}} \\
      \caption{The DLA detected towards QSO 1036-272 at $z_{\rm abs}=2.792$. The solid red line represents the Voigt profile fit to the DLA with an inferred column density of log $N_{{\rm H}\,{\sc \rm I}}=20.65\pm0.13$. The system is detected down to Ly$\beta$.}
         \label{Q1036-272}
   \end{figure}

    \begin{figure}
   \centering
      {\includegraphics[width=0.9\columnwidth,clip=]{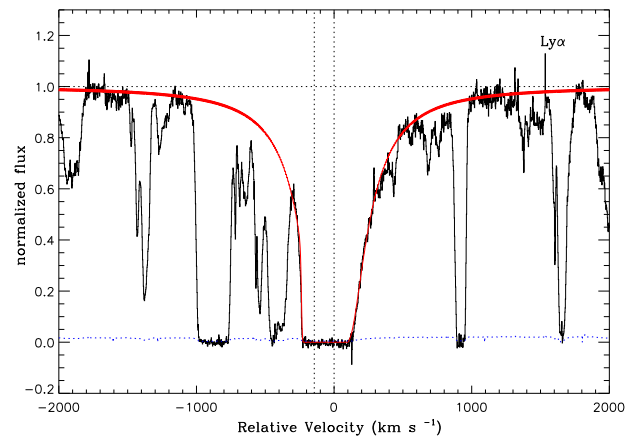}} \\
      \caption{The sub-DLA detected towards QSO 1036-2257 at $z_{\rm abs}=2.533$. The solid red line represents the Voigt profile fit to the sub-DLA with a total column density of log $N_{{\rm H}\,{\sc \rm I}}=19.30\pm0.10$ fitted with two velocity components at 0 and -144 km s$^{-1}$.}
         \label{Q1036-2257}
   \end{figure}

    \begin{figure}
   \centering
      {\includegraphics[width=\columnwidth,height=7.5cm,clip=]{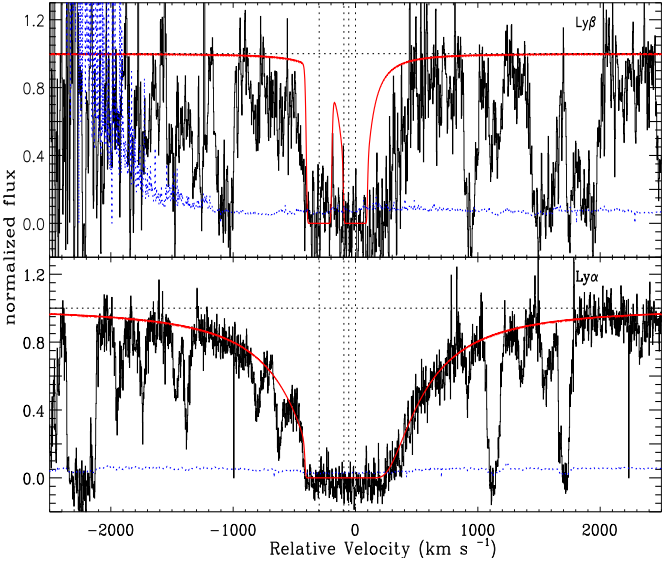}} \\
      \caption{The sub-DLA detected towards QSO B1036-268 at $z_{\rm abs}=2.235$. Ly$\beta$ region of the spectrum is smoothed with a boxcar average of 2.0 pixels for plotting purposes. The solid red line corresponds to the Voigt profile fit to the sub-DLA with a total column density of log $N_{{\rm H}\,{\sc \rm I}}=19.96\pm0.09$ from 0, -55, -95, and -297 km s$^{-1}$ velocity components. The system is detected down to Ly$\beta$.}
         \label{B1036-268}
   \end{figure}

\item QSO B0952-0115 ($z_{\rm em}=4.426$). One damped absorber was previously reported in this quasar at $z_{\rm abs}=4.024$ \citep{storrie00,prochaska07}. A Lyman limit system at $z_{\rm LLS}=4.242$ is detected in the spectrum. We find a sub-DLA at $z_{\rm abs}=3.476$ with log $N_{{\rm H}\,{\sc \rm I}}=20.04\pm0.07$ and $b=32\pm3.9$ km s$^{-1}$. Metal lines from {Fe}\,{\sc ii} $\lambda\lambda$ 1608, 1611, {Si}\,{\sc ii} $\lambda\lambda\lambda$ 1260, 1304, 1526, {Al}\,{\sc ii} $\lambda$ 1670, {Al}\,{\sc iii} $\lambda\lambda$ 1854, 1862, and {C}\,{\sc iv} $\lambda\lambda$ 1548, 1550 associated with the sub-DLA are covered in the red part of the spectrum. Fig. \ref{B0952-0115} shows the best fit result of the {H}\,{\sc i} column density.

\item Q1036-272 ($z_{\rm em}=3.090$). A low-resolution spectrum of this quasar has been previously published \citep{jakobsen92}. We report a DLA with $\ion{H}{i}$ column density of log $N_{{\rm H}\,{\sc \rm I}}=20.65\pm0.13$ and $b=35\pm5$ km s$^{-1}$ down to Ly$\beta$ at $z_{\rm abs}=2.792$. Several metal absorption lines from {O}\,{\sc i} $\lambda$ 1302, {Fe}\,{\sc ii} $\lambda\lambda\lambda\lambda$ 1144, 2344, 2374, 2382, {Si}\,{\sc ii} $\lambda\lambda\lambda\lambda$ 1190, 1193, 1260, 1304, and {Al}\,{\sc iii} $\lambda\lambda$ 1854, 1862 are covered in the red part of the spectrum. Fig. \ref{Q1036-272} shows best fit result of the {H}\,{\sc i} lines at $z_{\rm abs}=2.792$.

   \begin{figure}
   \centering
      {\includegraphics[width=0.9\columnwidth,clip=]{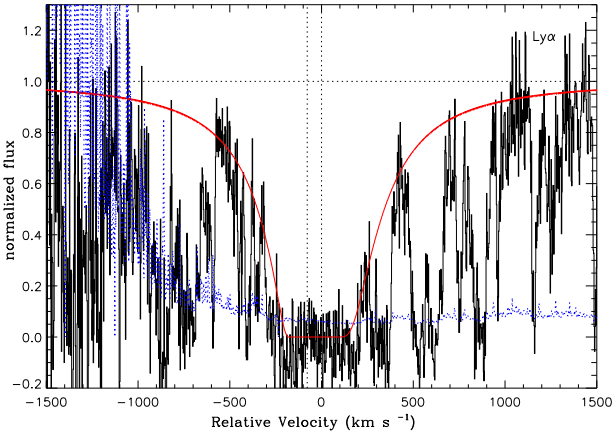}} \\
      \caption{The sub-DLA detected towards LBQS\,1232$+$0815 at $z_{\rm abs}=1.720$. The spectrum is smoothed with a boxcar average of 0.5 pixel for plotting purposes. The solid red line represents the Voigt profile fit to the sub-DLA with a total column density of log $N_{{\rm H}\,{\sc \rm I}}=19.48\pm0.13$ using 0 and -78 km s$^{-1}$ velocity components (see Fig. \ref{met1232}).}
         \label{1232+0815}
   \end{figure}

   \begin{figure}
  \centering
  {\includegraphics[width=\columnwidth,height=8.0cm,clip=]{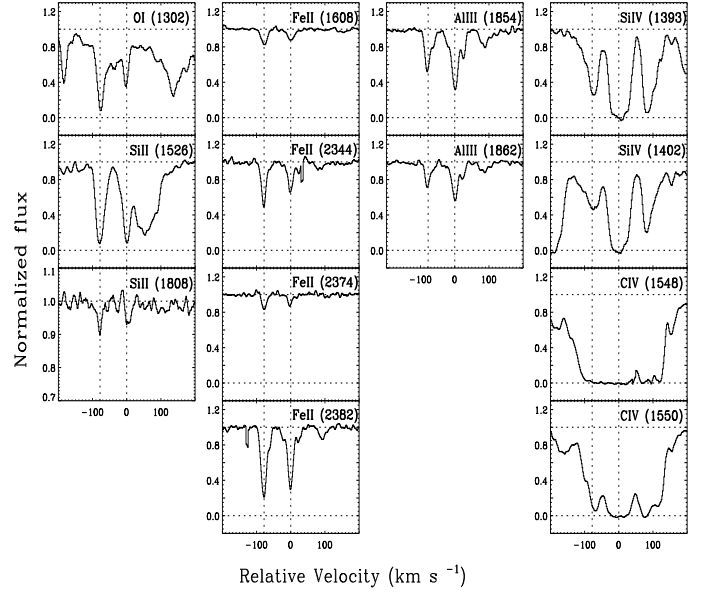}} \\
  \caption{Low and high ionization metal lines associated with the sub-DLA at $z_{\rm abs}=1.720$ (zero velocity) along the line-of-sight of LBQS 1232$+$0815 are shown. The line IDs are given in each panel.}
    \label{met1232}
   \end{figure}

\item QSO B1036-2257 ($z_{\rm em}=3.130$). A damped and sub-damped absorber were previously reported in this quasar at $z_{\rm abs}=2.777$ \citep{fox09} and $z_{\rm abs}=2.531$ \citep{lopez01} respectively. A Lyman limit system at $z_{\rm LLS}=2.792$ is also detected. From a low-resolution spectrum, \citet{lopez01} measured the equivalent width of the {H}\,{\sc i} absorption line to be ${\rm EW_{obs}}=13.52\,\AA$ but state that they cannot measure the {H}\,{\sc i} column density due to the limited spectral resolution. Using the high-resolution UVES spectrum, we are able to identify two main components in the system from the metal lines at 0 and -144 km s$^{-1}$. Fitting these components we measure the total column density of the sub-DLA to be log $N_{{\rm H}\,{\sc \rm I}}=19.30\pm0.10$ with $b$ parameter of $b=26\pm3.4$ and fixed 20 km s$^{-1}$ (at 0 and -144 km s$^{-1}$). The component at 0 km s$^{-1}$ is stronger and heavily dominates over the component at 144 km s$^{-1}$. Metal lines from {Fe}\,{\sc ii} $\lambda\lambda\lambda\lambda$ 1144, 2344, 2382, 2586, {Si}\,{\sc ii} $\lambda\lambda\lambda$ 1193, 1260, 1526, {C}\,{\sc ii} $\lambda$ 1334, {Al}\,{\sc iii} $\lambda\lambda$ 1854, 1862, {Si}\,{\sc iv} $\lambda\lambda$ 1393, 1402, and {C}\,{\sc iv} $\lambda\lambda$ 1548, 1550 associated with the sub-DLA are detected in the red part of the spectrum. Fig. \ref{Q1036-2257} shows the best fit result of the {H}\,{\sc i} column density.

  \begin{figure}
   \centering
      {\includegraphics[width=0.9\columnwidth,clip=]{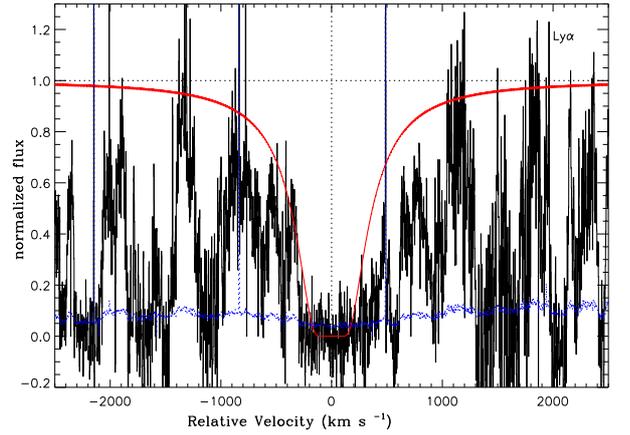}} \\
      \caption{The sub-DLA detected towards QSO J1330-2522 at $z_{\rm abs}=2.654$. The solid red line shows the Voigt profile fit to the sub-DLA with an inferred column density of log $N_{{\rm H}\,{\sc \rm I}}=19.56\pm0.13$.}
        \label{J1330-2522}
   \end{figure}

 \begin{figure}
 \centering
  {\includegraphics[width=\columnwidth,clip=]{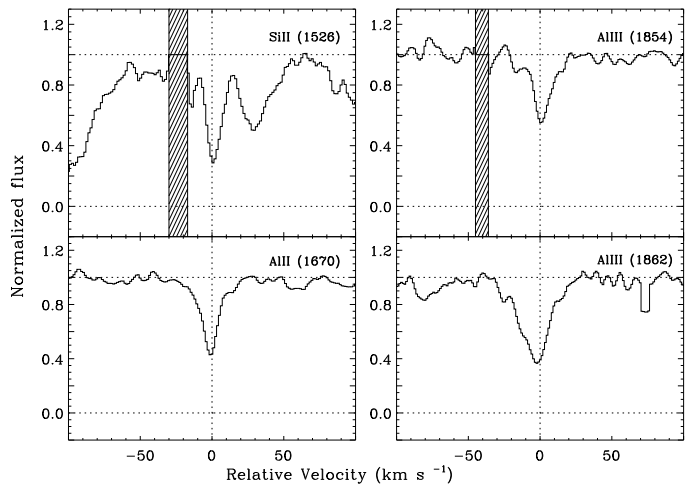}} \\
      \caption{Low ionization metal lines associated with the sub-DLA at $z_{\rm abs}=2.654$ along the line-of-sight of LBQS 1330-2522. The vertical dashed line is adopted zero velocity corresponding to $z_{\rm abs}=2.654$. The line IDs are given in each panel. The shaded area corresponds to the regions affected by cosmic rays.}
         \label{met1330}
   \end{figure}

   \begin{figure}
   \centering
      {\includegraphics[width=0.9\columnwidth,clip=]{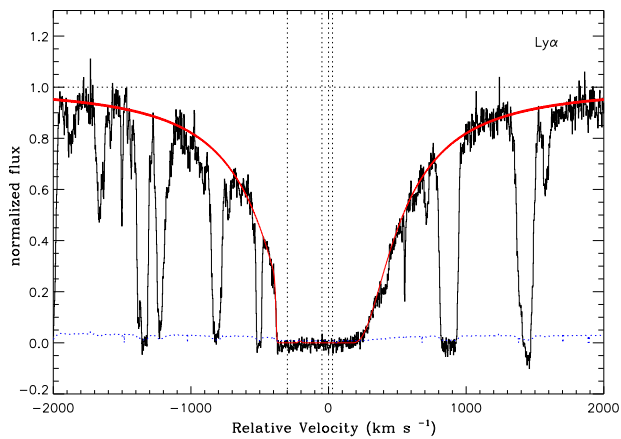}} \\
      \caption{The sub-DLA detected towards QSO J1356-1101 at $z_{\rm abs}=2.397$. The solid red line represents the Voigt profile fit to the sub-DLA with a total column density of log $N_{{\rm H}\,{\sc \rm I}}=19.88\pm0.09$ using 29, 0, -48 and -300 km s$^{-1}$ velocity components.}
         \label{J1356-1101}
   \end{figure}

\item QSO B1036-268 ($z_{\rm em}=2.460$). A low-resolution slit spectrum of the quasar has been previously published \citep{jakobsen92}. We find four strong velocity components in the system at $z_{\rm abs}=2.235$ from the metal lines. The column density of the sub-DLA fitted down to Ly$\beta$ with four components at 0, -55, -95, and -144 km s$^{-1}$, resulting a total of log $N_{{\rm H}\,{\sc \rm I}}=19.96\pm0.09$ and $b=20$, 20, 20 (fixed), and $35\pm3.0$ km s$^{-1}$ respectively. The component at 0 km s$^{-1}$ is strongest and heavily dominates over other components. Strong metal absorption lines from {O}\,{\sc i} $\lambda$ 1302, {Fe}\,{\sc ii} $\lambda\lambda\lambda\lambda\lambda\lambda\lambda$ 1144, 1608, 1611, 2344, 2374, 2382, 2586, {Si}\,{\sc ii} $\lambda\lambda\lambda\lambda\lambda$ 1190, 1193, 1260, 1304, 1526, {C}\,{\sc ii} $\lambda$ 1334, {Al}\,{\sc ii} $\lambda$ 1670, {Mg}\,{\sc ii} $\lambda\lambda$ 2796, 2803, {Al}\,{\sc iii} $\lambda\lambda$ 1854, 1862, {Si}\,{\sc iv} $\lambda\lambda$ 1393, 1402, and {C}\,{\sc iv} $\lambda\lambda$ 1548, 1550 at the redshift of the sub-DLA are covered in the red part of the spectrum. Fig. \ref{B1036-268} shows our best fit result of {H}\,{\sc i} lines of the sub-DLA.
 
\item LBQS 1232+0815 ($z_{\rm em}=2.570$). A DLA was known at $z_{\rm abs}=2.338$ along the line-of-sight of this quasar \citep{prochaska07,ivanchik10}. We report for the first time a sub-DLA at $z_{\rm abs}=1.720$ with total log $N_{{\rm H}\,{\sc \rm I}}=19.48\pm0.13$. From $\ion{O}{i}$, we identified two main velocity components from the system at 0 and -78 km s$^{-1}$ (see Fig. \ref{met1232}) where $b$ is fixed at $b=20$ km s$^{-1}$ for both components. Metal absorption lines from {O}\,{\sc i} $\lambda$ 1302, {Fe}\,{\sc ii} $\lambda\lambda\lambda\lambda\lambda$ 1608, 1611, 2344, 2374, 2382, {Si}\,{\sc ii} $\lambda\lambda$ 1526, 1808, {Al}\,{\sc iii} $\lambda\lambda$ 1854, 1862, {Si}\,{\sc iv} $\lambda\lambda$ 1393, 1402, and {C}\,{\sc iv} $\lambda\lambda$ 1548, 1550 associated with the sub-DLA are covered in the red part of the spectrum. Fig. \ref{1232+0815} shows the best fit result of {H}\,{\sc i} column density of the sub-DLA. While the Ly$\alpha$ line is noisy, the presence of the sub-DLA is confirmed through the metal lines detected in the spectrum (see Fig. \ref{met1232}).

   \begin{figure}
   \centering
      {\includegraphics[width=\columnwidth,height=9cm,clip=]{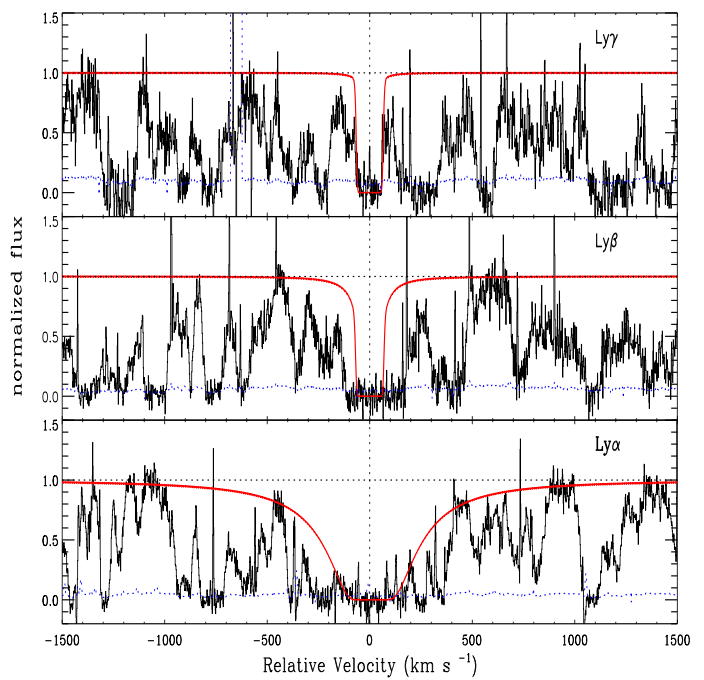}} \\
      \caption{The sub-DLA detected towards J1723$+$2243 at $z_{\rm abs}=4.155$. The solid red line corresponds to the Voigt profile fit to the sub-DLA with an inferred column density of log $N_{{\rm H}\,{\sc \rm I}}=19.23\pm0.12$. The system is detected down to the Ly$\gamma$ range.}
         \label{J1723+2243}
   \end{figure}

   \begin{figure}
   \centering
      {\includegraphics[width=0.9\columnwidth,clip=]{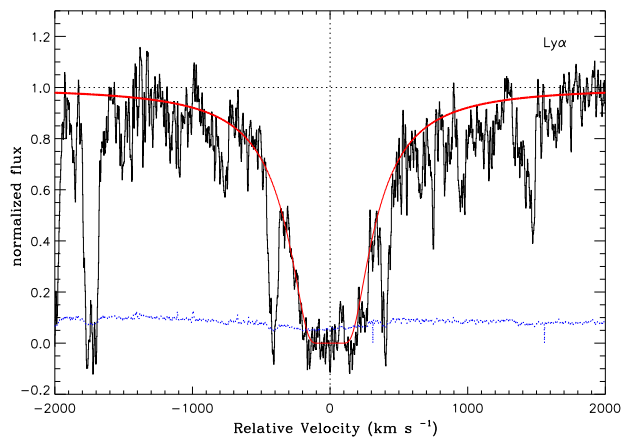}} \\
      \caption{The sub-DLA detected towards LBQS\,2114-4347 at $z_{\rm abs}=1.912$. For plotting purposes the spectrum is smoothed with a boxcar average of 1.0 pixel. The solid red line shows the Voigt profile fit to the sub-DLA with an inferred column density of log $N_{{\rm H}\,{\sc \rm I}}=19.50\pm0.10$.}
        \label{2114-4347}
   \end{figure}

\item QSO J1330-2522 ($z_{\rm em}=3.910$). Two sub-DLAs were previously reported at $z_{\rm abs}=2.910$ and 3.080 \citep{peroux01} along the line-of-sight of this quasar. A Lyman limit system at $z_{\rm LLS}=3.728$ is also seen. We report a new sub-DLA at $z_{\rm abs}=2.654$ with neutral hydrogen column density of log $N_{{\rm H}\,{\sc \rm I}}=19.56\pm0.13$ and $b=25.5\pm2.4$ km s$^{-1}$. Fig. \ref{J1330-2522} shows our best fit result. Metal lines from {Si}\,{\sc ii} $\lambda\lambda\lambda$ 1260, 1526, 1808, {Al}\,{\sc ii} $\lambda$ 1670, {Al}\,{\sc iii} $\lambda\lambda$ 1854, 1862, {Si}\,{\sc iv} $\lambda\lambda$ 1393, 1402, and {C}\,{\sc iv} $\lambda\lambda$ 1548, 1550 are covered in the red part of the spectrum at the redshift of the sub-DLA. As an example of metal lines, low ionization lines from the sub-DLA confirming its presence are plotted in Fig. \ref{met1330}. High ionization lines are blended with the lines from the other two sub-DLAs.

\item QSO J1356-1101 ($z_{\rm em}=3.006$). Two damped absorbers have been previously reported in this quasar at $z_{\rm abs}=2.501$ and 2.967 \citep{prochaska07,noterdaeme08,fox09}. We report for the first time a sub-DLA at $z_{\rm abs}=2.397$ in the quasar. From the $\ion{O}{i}$, we find four strong components in the system at 29, 0, -48, and -300 km s$^{-1}$. The total $\ion{H}{i}$ column density of the system is log $N_{{\rm H}\,{\sc \rm I}}=19.88\pm0.09$ with $b=20$, 20, 20, (fixed) $28\pm2.8$ km s$^{-1}$ for 29, 0, -48, and -300 km s$^{-1}$ components respectively. The component at 0 km s$^{-1}$ is strongest and heavily dominates over other components. Metal absorption lines from {O}\,{\sc i} $\lambda$ 1302, {Fe}\,{\sc ii} $\lambda\lambda\lambda\lambda\lambda$ 1144, 2344, 2374, 2382, 2586, and {Si}\,{\sc ii} $\lambda\lambda\lambda$ 1190, 1193, 1260, 1304 associated with the sub-DLA are detected in the red part of the spectrum. Fig. \ref{J1356-1101} shows the best fit result of the neutral hydrogen column density of the sub-DLA. 

    \begin{figure}
   \centering
      {\includegraphics[width=0.9\columnwidth,clip=]{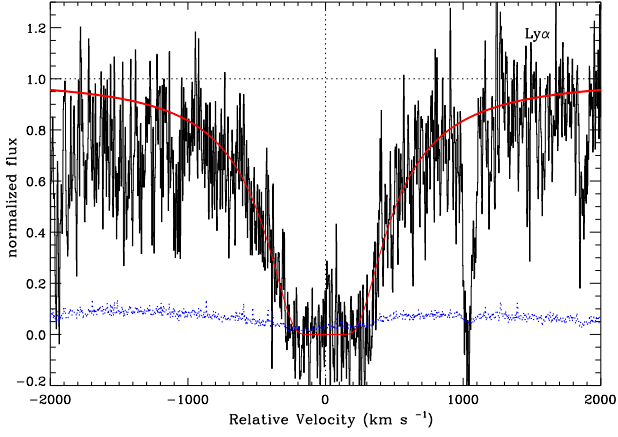}} \\
      \caption{The sub-DLA detected towards J223941.8-294955 at $z_{\rm abs}=1.825$. The spectrum is smoothed with a boxcar average of 1.0 pixel for plotting reasons. The solid red line corresponds to the Voigt profile fit to the sub-DLA with an inferred column density of log $N_{{\rm H}\,{\sc \rm I}}=19.84\pm0.14$. Emission is clearly seen in the trough of this absorber which is likely to correspond to the Lyman-$\alpha$ emission from the sub-DLA host.}
        \label{J223941.8-294955}
   \end{figure}

   \begin{figure}
   \centering
      {\includegraphics[width=0.9\columnwidth,clip=]{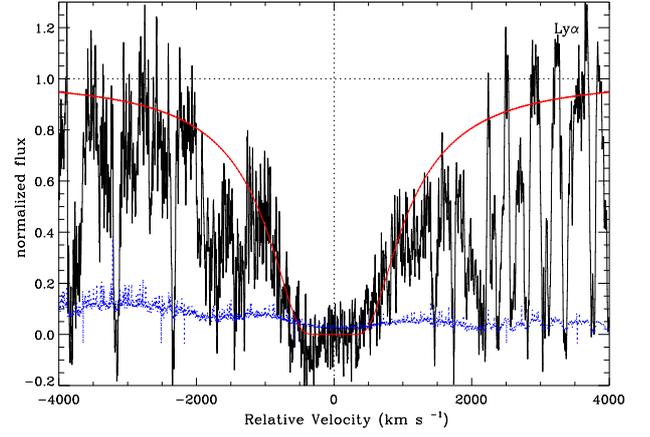}} \\
      \caption{The DLA detected towards QSO B2318-1107 at $z_{\rm abs}=1.629$. For plotting purposes the spectrum is smoothed with a boxcar average of 0.5 pixel. The solid red line corresponds to the Voigt profile fit to the DLA with an inferred column density of log $N_{{\rm H}\,{\sc \rm I}}=20.52\pm0.14$.}
        \label{B2318-1107}
   \end{figure}

\item QSO J1723+2243 ($z_{\rm em}=4.520$). A damped absorber has been previously reported in this quasar at $z_{\rm abs}=3.697$ \citep{prochaska07,guimares09}. We observed a sub-DLA down to Ly$\gamma$ at $z_{\rm abs}=4.155$ with {H}\,{\sc i} column of log $N_{{\rm H}\,{\sc \rm I}}=19.23\pm0.12$ with a fixed $b=20$ km s$^{-1}$. The metal lines associated with this system are either blended with other features or not covered by our data so that this detection is based on the Lyman line only and is a little less secure than the others. The detection of absorption lines from the Lyman series confirm the presence of the sub-DLA. Fig. \ref{J1723+2243} shows our best fit result of {H}\,{\sc i} lines.


   \begin{figure}
   \centering
      {\includegraphics[width=\columnwidth,height=9cm,clip=]{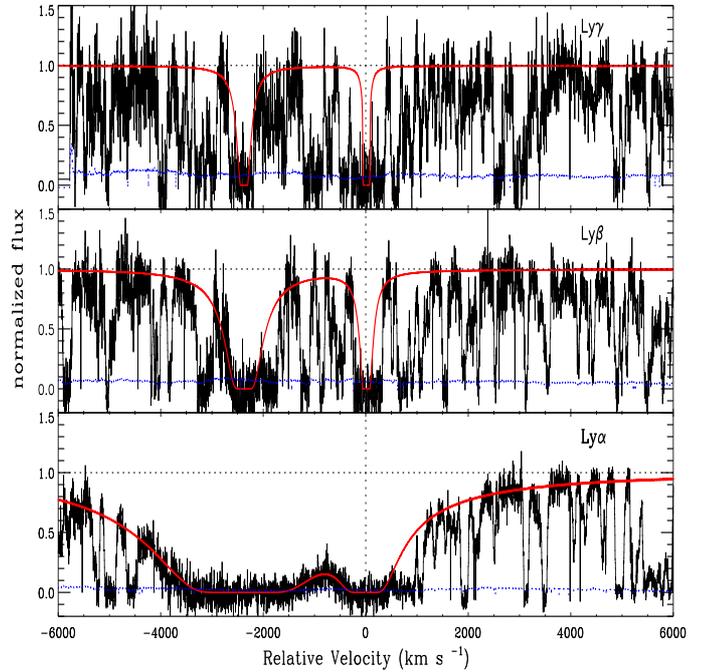}} \\
      \caption{The sub-DLA detected towards QSO B2342$+$3417 at $z_{\rm abs}=2.940$. The solid red line corresponds to the Voigt profile fit to the sub-DLA with an inferred column density of log $N_{{\rm H}\,{\sc \rm I}}=20.18\pm0.10$. The system is detected down to the Ly$\gamma$. The system is fitted simultaneously with a high column density absorber at $z_{\rm abs}=2.909$ (log $N_{{\rm H}\,{\sc \rm I}}=21.10\pm0.10$) to determine the column density precisely.}
         \label{B2342+3417}
   \end{figure}

\item LBQS 2114-4347 ($z_{\rm em}=2.040$). The quasar has been discovered as part of the Large Bright Quasar Survey (LBQS; \citealt{morris91}). No absorber has been previously reported in this quasar \citep{peroux03}. We observed, for the first time, a sub-DLA at $z_{\rm abs}=1.912$ with best fit column density of log $N_{{\rm H}\,{\sc \rm I}}=19.50\pm0.10$ and $b=31.7\pm3.4$ km s$^{-1}$. Several metal absorption lines from {O}\,{\sc i} $\lambda$ 1302, {Fe}\,{\sc ii} $\lambda\lambda\lambda\lambda\lambda\lambda$ 1144, 1608, 2344, 2374, 2382, 2586, {Si}\,{\sc ii} $\lambda\lambda\lambda\lambda\lambda$ 1193, 1260, 1304, 1526, 1808, {C}\,{\sc ii} $\lambda$ 1334, {Al}\,{\sc ii} $\lambda$ 1670, {Mg}\,{\sc ii} $\lambda\lambda$ 2796, 2803, {Si}\,{\sc iv} $\lambda\lambda$ 1393, 1402, and {C}\,{\sc iv} $\lambda\lambda$ 1548, 1550 at the redshift of the sub-DLA are covered in the red part of the spectrum. Fig. \ref{2114-4347} shows our best fit of {H}\,{\sc i} column density.
 
\item J223941.8-294955 ($z_{\rm em}=2.102$). This quasar was discovered during the course of the 2dF quasar redshift survey. The absorber at $z_{abs}=1.825$ in this quasar has been reported before by \citet{cappetta10} with a column density of log $N_{{\rm H}\,{\sc \rm I}}=20.60$ (where {H}\,{\sc i} fit was not shown). We fit the {H}\,{\sc i} of the absorber again and find that the absorber at $z_{\rm abs}=1.825$ fits well with {H}\,{\sc i} column density of log $N_{{\rm H}\,{\sc \rm I}}=19.84\pm0.14$ and $b=53.0\pm4.7$ km s$^{-1}$. Emission is clearly seen in the trough of this absorber which is likely to correspond to the Lyman-$\alpha$ emission from the sub-DLA host. Several metal absorption lines from {O}\,{\sc i} $\lambda$ 1302, {Fe}\,{\sc ii} $\lambda\lambda\lambda\lambda\lambda\lambda$ 1144, 1608, 2344, 2374, 2382, 2586, {Si}\,{\sc ii} $\lambda\lambda\lambda\lambda\lambda$ 1193, 1260, 1304, 1526, 1808, {C}\,{\sc ii} $\lambda$ 1334, {Al}\,{\sc ii} $\lambda$ 1670, {Mg}\,{\sc ii} $\lambda\lambda$ 2796, 2803, {Al}\,{\sc iii} $\lambda\lambda$ 1854, 1862, {Si}\,{\sc iv} $\lambda\lambda$ 1393, 1402, and {C}\,{\sc iv} $\lambda\lambda$ 1548, 1550 associated with sub-DLA are covered in the red part of the spectrum. Fig. \ref{J223941.8-294955} shows our best fit result of neutral hydrogen column density fit. 

\item QSO B2318-1107 ($z_{\rm em}= 2.960$). A DLA was previously known at $z_{\rm abs}=1.989$ along the line-of-sight of the quasar \citep{noterdaeme07,fox09}. We find a new DLA in the quasar at $z_{\rm abs}=1.629$ with neutral hydrogen column of log $N_{{\rm H}\,{\sc \rm I}}=20.52\pm0.14$ with fixed $b=30.0$ km s$^{-1}$. Several metal absorption lines from {Fe}\,{\sc ii} $\lambda\lambda\lambda\lambda\lambda$ 1608, 1611, 2344, 2374, 2382, {Si}\,{\sc ii} $\lambda\lambda\lambda\lambda$ 1193, 1260, 1304, 1526 {C}\,{\sc ii} $\lambda$ 1334, {Al}\,{\sc ii} $\lambda$ 1670, {Al}\,{\sc iii} $\lambda\lambda$ 1854, 1862, and {Si}\,{\sc iv} $\lambda\lambda$ 1393, 1402 associated with the DLA are covered in the red part of the spectrum. Fig. \ref{B2318-1107} shows our best fit of the {H}\,{\sc i} column to the DLA.


\item QSO B2342+3417 ($z_{\rm em}=3.010$). A damped absorber with log $N_{{\rm H}\,{\sc \rm I}}=21.10\pm0.10$ was previously reported at $z_{\rm abs}=2.909$ in the quasar \citep{prochaska03b,fox09}. A joint fit was implemented by \citet{prochaska03b} to fit the DLA together with the neighboring sub-DLA, but the column density of the sub-DLA was not reported. We measure log $N_{{\rm H}\,{\sc \rm I}}=20.18\pm0.10$ down to Ly$\gamma$ at $z_{\rm abs}=2.940$ with a fixed $b=20.0$ km s$^{-1}$. The metal lines associated with this system are not seen because of limited wavelength coverage in the red part of the spectrum. Fig. \ref{B2342+3417} shows the best fit result of the {H}\,{\sc i} lines.
\end{enumerate}

\subsection{DLAs/sub-DLAs towards EUADP quasars}
Besides the 150 DLAs/sub-DLAs, we found another 47 damped absorbers (21 DLAs and 26 sub-DLAs) in the literature along the lines-of-sight of our 250 EUADP quasars, for which Ly$\alpha$ absorption lines are not covered by our data either due to the limited wavelength coverage or non-overlapping settings. These systems are however of interest to us because their metal lines might still be included in our data and are helpful in further studies of the EUADP sample. These 150 and 47 damped absorbers (with and without Ly$\alpha$ covered by the EUADP dataset) make up a total of 197 DLAs/sub-DLAs along lines-of-sight of the 250 EUADP quasars where 114 are DLAs and 83 are sub-DLAs. The EUADP sample by design is biased towards DLAs and therefore we see less sub-DLAs than DLAs in the sample. Indeed, in the redshift range 0.2 $<$ $z$ $<$ 4.9, we expect twice as many sub-DLAs as DLAs based on the number density of absorbers at a mean redshift of $z=2.4$ \citep[see][]{peroux05}.

\section{Global Properties of the EUADP sample}
  \begin{figure}
   \centering
      {\includegraphics[width=\columnwidth,clip=]{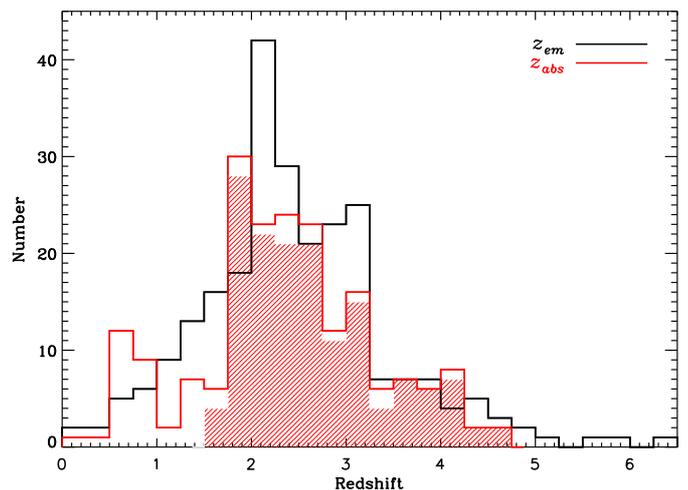}} \\
      \caption{Distribution of quasar emission redshift of the EUADP sample (black histogram). The red histogram shows the redshift distribution of DLAs/sub-DLAs along the lines-of-sight of EUADP quasars. The red shaded area corresponds to the distribution of the 150 damped absorbers with Ly$\alpha$ covered by our EUADP sample.}
               \label{zemhist}
   \end{figure}

The emission redshifts of the 250 EUADP sample quasars are initially obtained from the Simbad catalogue and later double checked for the cases where Ly$\alpha$ emission from the quasar is covered by our data. The Ly$\alpha$ emission for 5 quasars (i.e., QSO\,J0332-4455, QSO\,B0528-2505, QSO\,B0841+129, QSO\,B1114-0822 and QSO\,J2346+1247) is not seen because of the presence of DLAs belonging to the ``proximate DLA'' class with $z_{\rm abs}\approx z_{\rm em}$ \citep[e.g.,][]{moller98,ellison11,zafar11}. For the emission redshifts of these 5 cases, we rely on the literature. The emission redshifts of all the other objects in the EUADP sample have been compared with measurements from the literature. For a few cases, emission redshifts provided in the Simbad catalogue are not correct and the correct redshifts are obtained from the literature. In our sample, there are 38 quasars with emission redshifts below $z_{\rm em}<1.5$. For these cases we cannot see Ly$\alpha$ emission from the quasar because of the limited spectral coverage, therefore, we relied mostly on the Simbad catalogue. However, other emission lines are covered in the spectra, helping us to confirm the emission redshifts. The emission redshifts of 250 quasars of EUADP sample ranges from $0.191\leq z_{\rm em}\leq6.311$. Their distribution is shown in Fig. \ref{zemhist} and is found to peak at $z_{\rm em}\simeq2.1$.

   \begin{figure}
   \centering
      {\includegraphics[width=\columnwidth,clip=]{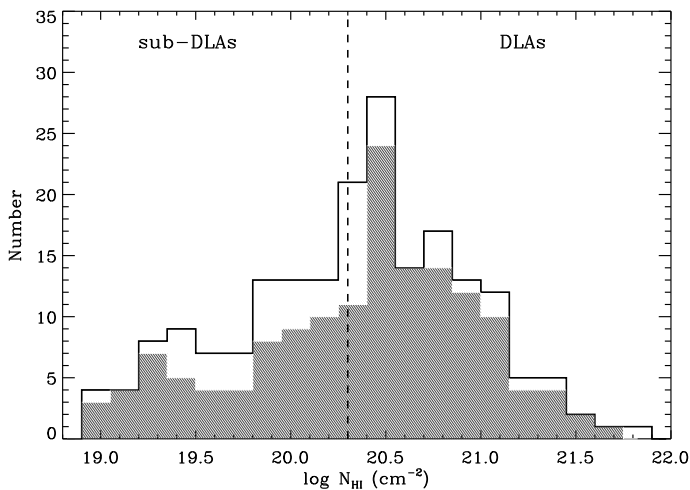}} \\
      \caption{Histogram showing the number of DLAs and sub-DLAs along lines-of-sight to the quasars in the EUADP sample. The shaded area represents the column density distribution of damped absorbers for which the Ly$\alpha$ line is covered by the EUADP data. The vertical dashed line is the dividing line between the sub-DLAs and DLAs classes.}
         \label{zabshist}
   \end{figure} 

In Fig. \ref{zabshist}, the column density distribution of DLAs and sub-DLAs is presented (see \citealt{zafar12b} for complete list of DLAs and sub-DLAs {H}\,{\sc i} column densities). It is worth noting that in the EUADP sample, damped absorbers with column densities up to log $N_{{\rm H}\,{\sc \rm I}}=21.85$ are seen, while higher column densities have been recently reported \citep{guimaraes12,kulkarni12,noterdaeme12}. As mentioned above in the EUADP sample, the number of sub-DLAs is lower than the number of DLAs. Indeed, a large fraction ($\sim$45\%) of the quasars in the EUADP sample were observed because of a known strong damped absorber along their line-of-sight. A carefully selected subset of the EUADP will have to be built for the purpose of statistical analysis of DLAs and sub-DLAs \citep{zafar12b}.


\section{Lines-of-sight of Interest}
In the EUADP sample, a few lines-of-sight of quasars are rich and contain more than one absorber. One interesting example is the line-of-sight of QSO J0133$+$0400, containing six DLAs and sub-DLAs. In this line-of-sight two sub-DLAs $z_{\rm abs}=3.995/3.999$ are separated by only $\Delta v=240$ km s$^{-1}$. Three more examples of rich lines-of-sight are: $i)$ QSO J0006-6208 with 3 DLAs and one sub-DLA, $ii)$ QSO J0407-4410 with 4 DLAs, where three DLAs have log $N_{{\rm H}\,{\sc \rm I}}\gtrsim21.0$, and $iv)$ QSO B0841$+$129 containing 3 DLAs (with log $N_{{\rm H}\,{\sc \rm I}}\gtrsim21.0$) and one sub-DLA. Such complex group or systems can be classified as multiple DLAs (MDLAs; \citealt{lopez03}).

In addition, there are four quasar pairs in the EUADP sample: $i)$ QSO J0008-2900 \& QSO J0008-2901 separated by $1.3'$. Two sub-DLAs at  $z=2.254$ and $z=2.491$ are seen along the lines of sight of QSO J0008-2900 and QSO J0008-2901 respectively. $ii)$ J030640.8-301032 \& J030643.7-301107 separated by $\sim$0.85 arcmin and no absorber is seen along the line-of-sight to the pair. $iii)$ QSO B0307-195A \& B separated by $\sim$1 arcmin. A sub-DLA (at $z=1.788$ with log $N_{{\rm H}\,{\sc \rm I}}=19.00\pm0.10$) along the line-of-sight of QSO B0307-195B is detected \citep{odorico02}, but is not seen in its companion. $iv)$ QSO J1039-2719 \& QSO B1038-2712 separated by $17.9'$. A sub-DLA (at $z=2.139$ with log $N_{{\rm H}\,{\sc \rm I}}=19.90\pm0.05$) along the line-of-sight of QSO J1039-2719 is detected \citep{odorico02}, but no damped absorber is seen in its companion.

\section{Conclusion}
In this study, high-resolution spectra taken from the UVES Advanced Data Products archive have been processed and combined to make a sample of 250 individual quasars spectra. The high-resolution spectra of these quasars allow us to detect absorbers down to log $N_{{\rm H}\,{\sc \rm I}}=19.00$ cm$^{-2}$. Automated and visual searches for quasar absorbers have been undertaken leading to a sample of 93 DLAs and 57 sub-DLAs. An extensive search in the literature shows that 6 of these DLAs and 13 of these sub-DLAs have their {H}\,{\sc i} column densities measured for the first time, where 10 are new identifications. These new damped absorbers are confirmed by detecting metal lines associated with the absorber and/or lines from the higher members of the Lyman series. The {H}\,{\sc i} column densities of all these new absorbers are determined by fitting a Voigt profile to the Ly$\alpha$ line together with the lines from higher order of the Lyman series whenever covered. Our data contain five proximate DLA cases and three quasar pairs. We found that a few lines-of-sight of quasars are very rich, particularly the line-of-sight of QSO J0133+0400 which contains six DLAs and sub-DLAs. 

In an accompanying paper, \citep{zafar12b}, we use a carefully selected subset of this dataset to study the statistical properties of DLAs and sub-DLAs, measure their column density distribution, and quantify the contribution of sub-DLAs to the {H}\,{\sc i} gas mass density. Further studies using specifically designed subsets of the EUADP dataset will follow.

\section{Acknowledgments}
This work has been funded within the BINGO! (`history of Baryons: INtergalactic medium/Galaxies cO-evolution') project by the Agence Nationale de la Recherche (ANR) under the allocation ANR-08-BLAN-0316-01. We would like to thank the ESO staff for making the UVES Advanced Data Products available to the community.  We are thankful to Stephan Frank, Jean-Michel Deharveng and Bruno Milliard for helpful comments.


\bibliographystyle{mnras}
\bibliography{dla.bib}

\begin{thebibliography}{55}
\expandafter\ifx\csname natexlab\endcsname\relax\def\natexlab#1{#1}\fi

\bibitem[{Bouch{\'e}} et~al.(2007){Bouch{\'e}}, {Lehnert}, {Aguirre},
  {P{\'e}roux} \& {Bergeron}]{bouche07}
{Bouch{\'e}} N., {Lehnert} M.~D., {Aguirre} A., {P{\'e}roux} C., {Bergeron} J.,
  2007, \mnras, 378, 525

\bibitem[{Bouch{\'e}} et~al.(2005){Bouch{\'e}}, {Lehnert} \&
  {P{\'e}roux}]{bouche05}
{Bouch{\'e}} N., {Lehnert} M.~D., {P{\'e}roux} C., 2005, \mnras, 364, 319

\bibitem[{Bouch{\'e}} et~al.(2006){Bouch{\'e}}, {Lehnert} \&
  {P{\'e}roux}]{bouche06}
{Bouch{\'e}} N., {Lehnert} M.~D., {P{\'e}roux} C., 2006, \mnras, 367, L16

\bibitem[{Cappetta} et~al.(2010){Cappetta}, {D'Odorico}, {Cristiani}, {Saitta}
  \& {Viel}]{cappetta10}
{Cappetta} M., {D'Odorico} V., {Cristiani} S., {Saitta} F., {Viel} M., 2010,
  \mnras, 407, 1290

\bibitem[{Croom} et~al.(2001){Croom}, {Smith}, {Boyle} et~al.]{croom01}
{Croom} S.~M., {Smith} R.~J., {Boyle} B.~J., et~al., 2001, \mnras, 322, L29

\bibitem[{Dekker} et~al.(2000){Dekker}, {D'Odorico}, {Kaufer}, {Delabre} \&
  {Kotzlowski}]{dekker00}
{Dekker} H., {D'Odorico} S., {Kaufer} A., {Delabre} B., {Kotzlowski} H., 2000,
  in { Society of Photo-Optical Instrumentation Engineers (SPIE) Conference
  Series\/}, edited by {M.~Iye \& A.~F.~Moorwood}, vol. 4008 of { SPIE
  Conference Series\/},  534--545

\bibitem[{Dessauges-Zavadsky} et~al.(2003){Dessauges-Zavadsky}, {P{\'e}roux},
  {Kim}, {D'Odorico} \& {McMahon}]{dessauges03}
{Dessauges-Zavadsky} M., {P{\'e}roux} C., {Kim} T.-S., {D'Odorico} S.,
  {McMahon} R.~G., 2003, \mnras, 345, 447

\bibitem[{D'Odorico} et~al.(2002){D'Odorico}, {Petitjean} \&
  {Cristiani}]{odorico02}
{D'Odorico} V., {Petitjean} P., {Cristiani} S., 2002, \aap, 390, 13

\bibitem[{Ellison} et~al.(2011){Ellison}, {Prochaska} \& {Mendel}]{ellison11}
{Ellison} S.~L., {Prochaska} J.~X., {Mendel} J.~T., 2011, \mnras, 412, 448

\bibitem[{Fontana} \& {Ballester}(1995)]{fontana}
{Fontana} A., {Ballester} P., 1995, The Messenger, 80, 37

\bibitem[{Fox} et~al.(2009){Fox}, {Prochaska}, {Ledoux}, {Petitjean}, {Wolfe}
  \& {Srianand}]{fox09}
{Fox} A.~J., {Prochaska} J.~X., {Ledoux} C., {Petitjean} P., {Wolfe} A.~M.,
  {Srianand} R., 2009, \aap, 503, 731

\bibitem[{Guimar{\~a}es} et~al.(2012){Guimar{\~a}es}, {Noterdaeme}, {Petitjean}
  et~al.]{guimaraes12}
{Guimar{\~a}es} R., {Noterdaeme} P., {Petitjean} P., et~al., 2012, \aj, 143,
  147

\bibitem[{Guimar{\~a}es} et~al.(2009){Guimar{\~a}es}, {Petitjean}, {de
  Carvalho} et~al.]{guimares09}
{Guimar{\~a}es} R., {Petitjean} P., {de Carvalho} R.~R., et~al., 2009, \aap,
  508, 133

\bibitem[{Ivanchik} et~al.(2010){Ivanchik}, {Petitjean}, {Balashev}
  et~al.]{ivanchik10}
{Ivanchik} A.~V., {Petitjean} P., {Balashev} S.~A., et~al., 2010, \mnras, 404,
  1583

\bibitem[{Jakobsen} \& {Perryman}(1992)]{jakobsen92}
{Jakobsen} P., {Perryman} M.~A.~C., 1992, \apj, 392, 432

\bibitem[{Kulkarni} et~al.(2007){Kulkarni}, {Khare}, {P{\'e}roux}, {York},
  {Lauroesch} \& {Meiring}]{kulkarni07}
{Kulkarni} V.~P., {Khare} P., {P{\'e}roux} C., {York} D.~G., {Lauroesch} J.~T.,
  {Meiring} J.~D., 2007, \apj, 661, 88

\bibitem[{Kulkarni} et~al.(2012){Kulkarni}, {Meiring}, {Som}
  et~al.]{kulkarni12}
{Kulkarni} V.~P., {Meiring} J., {Som} D., et~al., 2012, \apj, 749, 176

\bibitem[{Lamontagne} et~al.(2000){Lamontagne}, {Demers}, {Wesemael},
  {Fontaine} \& {Irwin}]{lamontagne00}
{Lamontagne} R., {Demers} S., {Wesemael} F., {Fontaine} G., {Irwin} M.~J.,
  2000, \aj, 119, 241

\bibitem[{Lanzetta} et~al.(1995){Lanzetta}, {Wolfe} \& {Turnshek}]{lanzetta95}
{Lanzetta} K.~M., {Wolfe} A.~M., {Turnshek} D.~A., 1995, \apj, 440, 435

\bibitem[{Lanzetta} et~al.(1991){Lanzetta}, {Wolfe}, {Turnshek}, {Lu},
  {McMahon} \& {Hazard}]{lanzetta91}
{Lanzetta} K.~M., {Wolfe} A.~M., {Turnshek} D.~A., {Lu} L., {McMahon} R.~G.,
  {Hazard} C., 1991, \apjs, 77, 1

\bibitem[{Ledoux} et~al.(2003){Ledoux}, {Petitjean} \& {Srianand}]{ledoux03}
{Ledoux} C., {Petitjean} P., {Srianand} R., 2003, \mnras, 346, 209

\bibitem[{Lopez} et~al.(2007){Lopez}, {Ellison}, {D'Odorico} \& {Kim}]{lopez07}
{Lopez} S., {Ellison} S., {D'Odorico} S., {Kim} T.-S., 2007, \aap, 469, 61

\bibitem[{Lopez} \& {Ellison}(2003)]{lopez03}
{Lopez} S., {Ellison} S.~L., 2003, \aap, 403, 573

\bibitem[{Lopez} et~al.(2001){Lopez}, {Maza}, {Masegosa} \& {Marquez}]{lopez01}
{Lopez} S., {Maza} J., {Masegosa} J., {Marquez} I., 2001, \aap, 366, 387

\bibitem[{Maza} et~al.(1996){Maza}, {Wischnjewsky} \& {Antezana}]{maza96}
{Maza} J., {Wischnjewsky} M., {Antezana} R., 1996, RMxAA, 32, 35

\bibitem[{Maza} et~al.(1995){Maza}, {Wischnjewsky}, {Antezana} \&
  {Gonz{\'a}lez}]{maza95}
{Maza} J., {Wischnjewsky} M., {Antezana} R., {Gonz{\'a}lez} L.~E., 1995, RMxAA,
  31, 119

\bibitem[{Meiring} et~al.(2008){Meiring}, {Kulkarni}, {Lauroesch}
  et~al.]{meiring08}
{Meiring} J.~D., {Kulkarni} V.~P., {Lauroesch} J.~T., et~al., 2008, \mnras,
  384, 1015

\bibitem[{Meiring} et~al.(2009){Meiring}, {Lauroesch}, {Kulkarni},
  {P{\'e}roux}, {Khare} \& {York}]{meiring09}
{Meiring} J.~D., {Lauroesch} J.~T., {Kulkarni} V.~P., {P{\'e}roux} C., {Khare}
  P., {York} D.~G., 2009, \mnras, 397, 2037

\bibitem[{M\o ller} et~al.(1998){M\o ller}, {Warren} \& {Fynbo}]{moller98}
{M\o ller} P., {Warren} S.~J., {Fynbo} J.~U., 1998, \aap, 330, 19

\bibitem[{Morris} et~al.(1991){Morris}, {Weymann}, {Anderson} et~al.]{morris91}
{Morris} S.~L., {Weymann} R.~J., {Anderson} S.~F., et~al., 1991, \aj, 102, 1627

\bibitem[{Morton}(2003)]{morton2003}
{Morton} D.~C., 2003, \apjs, 149, 205

\bibitem[{Noterdaeme} et~al.(2012{\natexlab{a}}){Noterdaeme}, {Laursen},
  {Petitjean} et~al.]{noterdaeme12}
{Noterdaeme} P., {Laursen} P., {Petitjean} P., et~al., 2012{\natexlab{a}},
  \aap, 540, A63

\bibitem[{Noterdaeme} et~al.(2007){Noterdaeme}, {Ledoux}, {Petitjean}, {Le
  Petit}, {Srianand} \& {Smette}]{noterdaeme07}
{Noterdaeme} P., {Ledoux} C., {Petitjean} P., {Le Petit} F., {Srianand} R.,
  {Smette} A., 2007, \aap, 474, 393

\bibitem[{Noterdaeme} et~al.(2008){Noterdaeme}, {Ledoux}, {Petitjean} \&
  {Srianand}]{noterdaeme08}
{Noterdaeme} P., {Ledoux} C., {Petitjean} P., {Srianand} R., 2008, \aap, 481,
  327

\bibitem[{Noterdaeme} et~al.(2012{\natexlab{b}}){Noterdaeme}, {Petitjean},
  {Carithers} et~al.]{noterdaeme12b}
{Noterdaeme} P., {Petitjean} P., {Carithers} W.~C., et~al., 2012{\natexlab{b}},
  \aap, 547, L1

\bibitem[{Noterdaeme} et~al.(2009){Noterdaeme}, {Petitjean}, {Ledoux} \&
  {Srianand}]{noterdaeme09}
{Noterdaeme} P., {Petitjean} P., {Ledoux} C., {Srianand} R., 2009, \aap, 505,
  1087

\bibitem[{Pagel}(2002)]{pagel02}
{Pagel} B.~E.~J., 2002, in { Chemical Enrichment of Intracluster and
  Intergalactic Medium\/}, edited by R.~{Fusco-Femiano}, F.~{Matteucci}, vol.
  253 of { Astronomical Society of the Pacific Conference Series\/},  489

\bibitem[{P{\'e}roux} et~al.(2003{\natexlab{a}}){P{\'e}roux},
  {Dessauges-Zavadsky}, {D'Odorico}, {Kim} \& {McMahon}]{peroux03b}
{P{\'e}roux} C., {Dessauges-Zavadsky} M., {D'Odorico} S., {Kim} T.-S.,
  {McMahon} R.~G., 2003{\natexlab{a}}, \mnras, 345, 480

\bibitem[{P{\'e}roux} et~al.(2005){P{\'e}roux}, {Dessauges-Zavadsky},
  {D'Odorico}, {Sun Kim} \& {McMahon}]{peroux05}
{P{\'e}roux} C., {Dessauges-Zavadsky} M., {D'Odorico} S., {Sun Kim} T.,
  {McMahon} R.~G., 2005, \mnras, 363, 479

\bibitem[{P{\'e}roux} et~al.(2003{\natexlab{b}}){P{\'e}roux}, {McMahon},
  {Storrie-Lombardi} \& {Irwin}]{peroux03}
{P{\'e}roux} C., {McMahon} R.~G., {Storrie-Lombardi} L.~J., {Irwin} M.~J.,
  2003{\natexlab{b}}, \mnras, 346, 1103

\bibitem[{P{\'e}roux} et~al.(2001){P{\'e}roux}, {Storrie-Lombardi}, {McMahon},
  {Irwin} \& {Hook}]{peroux01}
{P{\'e}roux} C., {Storrie-Lombardi} L.~J., {McMahon} R.~G., {Irwin} M., {Hook}
  I.~M., 2001, \aj, 121, 1799

\bibitem[{Pettini}(2004)]{pettini04}
{Pettini} M., 2004, in { Cosmochemistry. The melting pot of the elements\/},
  edited by C.~{Esteban}, R.~{Garc{\'{\i}}a L{\'o}pez}, A.~{Herrero},
  F.~{S{\'a}nchez},  257--298

\bibitem[{Pettini}(2006)]{pettini06}
{Pettini} M., 2006, in { The Fabulous Destiny of Galaxies: Bridging Past and
  Present\/}, edited by V.~{Le Brun}, A.~{Mazure}, S.~{Arnouts},
  D.~{Burgarella},  319

\bibitem[{Pettini} et~al.(1999){Pettini}, {Ellison}, {Steidel} \&
  {Bowen}]{pettini99}
{Pettini} M., {Ellison} S.~L., {Steidel} C.~C., {Bowen} D.~V., 1999, \apj, 510,
  576

\bibitem[{Prochaska} et~al.(2003){Prochaska}, {Gawiser}, {Wolfe}, {Cooke} \&
  {Gelino}]{prochaska03b}
{Prochaska} J.~X., {Gawiser} E., {Wolfe} A.~M., {Cooke} J., {Gelino} D., 2003,
  \apjs, 147, 227

\bibitem[{Prochaska} et~al.(2008){Prochaska}, {Hennawi} \&
  {Herbert-Fort}]{prochaska08}
{Prochaska} J.~X., {Hennawi} J.~F., {Herbert-Fort} S., 2008, \apj, 675, 1002

\bibitem[{Prochaska} et~al.(2005){Prochaska}, {Herbert-Fort} \&
  {Wolfe}]{prochaska05}
{Prochaska} J.~X., {Herbert-Fort} S., {Wolfe} A.~M., 2005, \apj, 635, 123

\bibitem[{Prochaska} et~al.(2007){Prochaska}, {Wolfe}, {Howk}, {Gawiser},
  {Burles} \& {Cooke}]{prochaska07}
{Prochaska} J.~X., {Wolfe} A.~M., {Howk} J.~C., {Gawiser} E., {Burles} S.~M.,
  {Cooke} J., 2007, \apjs, 171, 29

\bibitem[{Storrie-Lombardi} \& {Wolfe}(2000)]{storrie00}
{Storrie-Lombardi} L.~J., {Wolfe} A.~M., 2000, \apj, 543, 552

\bibitem[{Tytler} et~al.(2009){Tytler}, {Gleed}, {Melis} et~al.]{tytler09}
{Tytler} D., {Gleed} M., {Melis} C., et~al., 2009, \mnras, 392, 1539

\bibitem[{Wolfe} et~al.(2005){Wolfe}, {Gawiser} \& {Prochaska}]{wolfe05}
{Wolfe} A.~M., {Gawiser} E., {Prochaska} J.~X., 2005, \araa, 43, 861

\bibitem[{Wolfe} et~al.(1995){Wolfe}, {Lanzetta}, {Foltz} \&
  {Chaffee}]{wolfe95}
{Wolfe} A.~M., {Lanzetta} K.~M., {Foltz} C.~B., {Chaffee} F.~H., 1995, \apj,
  454, 698

\bibitem[{Wolfe} et~al.(2003){Wolfe}, {Prochaska} \& {Gawiser}]{wolfe03}
{Wolfe} A.~M., {Prochaska} J.~X., {Gawiser} E., 2003, \apj, 593, 215

\bibitem[{Zafar} et~al.(2011){Zafar}, {M{\o}ller}, {Ledoux} et~al.]{zafar11}
{Zafar} T., {M{\o}ller} P., {Ledoux} C., et~al., 2011, \aap, 532, A51

\bibitem[{Zafar} et~al.(2013){Zafar}, {P{\'e}roux}, {Popping}, {Milliard},
  {Deharveng} \& {Frank}]{zafar12b}
{Zafar} T., {P{\'e}roux} C., {Popping} A., {Milliard} B., {Deharveng} J.-M.,
  {Frank} S., 2013, \aap, 556, A141

\end{thebibliography}

\addtocounter{table}{-3}
\begin{table*}
\caption{Properties of the EUADP quasar sample. The details of the sample are provided in the columns as the (1) Simbad quasar names, (2) right ascension (RA), (3) declination (Dec), (4) Simbad $V$-band magnitude of each quasar except when noted otherwise, (5) emission redshift of the quasars, (6) wavelength coverage of the spectra, (7), ESO program ID, (8) number of spectra used for co-adding, where spectra from BLUE, RED upper and lower arms are counted separately, and (9) total exposure time of each quasar.}
\label{proptab}
\centering
\setlength{\tabcolsep}{1pt}
\begin{tabular}{@{} l c c c c c c c c @{}}
\hline
 Quasar & RA & Dec &  Mag. & $z_{\rm em}$ &  Wavelength coverage &  Prog. ID &  No. &  T$_{\rm exp}$ \\
 &  2000 & 2000 &  & & \AA\ &  & spec & sec  \\
\hline
LBQS 2359-0216B & 00 01 50.00 & -01 59 40.00 & 18.00 & 2.810 & 3290-5760, 5835-8520, 8660-10420 & 66.A-0624(A), 073.B-0787(A)Ê& 16 & 21,300 \\
QSO J0003-2323 &  00 03 44.92 & -23 23 54.80 & 16.70 & 2.280 & 3050-5760, 5835-8520, 8660-10420 & 166.A-0106(A)Ê& 36 & 43,200  \\
QSO B0002-422   & 00 04 48.28 & -41 57 28.10 & 17.40 & 2.760 & 3050-5760, 5835-8520, 8660-10420 & 166.A-0106(A) & 84 &  97,421  \\
QSO J0006-6208    & 00 06 51.61 &  -62 08 03.70 & 18.29 &  4.455 & 4780-5750, 5830-6800 & 69.A-0613(A)Ê & 4 & 7,200  \\
QSO J0008-0958    & 00 08 15.33 & -09 58 54.45 & 18.85 & 1.950 & 3300-4510,5720-7518,7690-9450 & 076.A-0376(A)Ê& 18 &  21,600 \\
QSO J0008-2900    & 00 08 52.72 & -29 00 43.75 & 19.12$^\dagger$ & 2.645 & 3300-4970,5730-10420 & 70.A-0031(A)Ê, 075.A-0617(A)Ê& 17 &  24,000  \\
QSO J0008-2901 & 00 08 57.74 & -29 01 26.61 & 19.85$^\dagger$ & 2.607 & 3300-4970,5730-10420 & 70.A-0031(A)Ê, 075.A-0617(A)Ê& 18 & 17,700 \\
QSO B0008+006  & 00 11 30.56 & +00 55 50.71 & 18.50 & 2.309 & 3760-4980,6700-8510,8660-10420 & 267.B-5698(A)Ê& 6 &  7,200 \\
LBQS 0009-0138    & 00 12 10.89 & -01 22 07.76 & 18.10 & 1.998 & 3300-4510,4770-5755,5835-6805 & 078.A-0003(A)Ê & 6 & 6,000 \\
LBQS 0010-0012    & 00 13 06.15 & +00 04 31.90  & 19.43 & 2.145 & 3050-3870,4610-5560,5670-6650 & 68.A-0600(A) & 3 & 5,400  \\
LBQS 0013-0029    & 00 16 02.41 & -00 12 25.08 & 17.00 & 2.087 & 3054-6800 & 66.A-0624(A)Ê, 267.A-5714(A)Ê & 66 & 88,200  \\
LBQS 0018+0026    & 00 21 33.28 & +00 43 00.99 & 18.20 & 1.244 & 3320-4510,4620-5600,5675-6650 & 078.A-0003(A)Ê& 3 &  3,000 \\
QSO B0027-1836  & 00 30 23.63 & -18 19 56.00  & 17.90 & 2.550 & 3050-3870,4780-5755,5835-6800 & 073.A-0071(A)Ê& 15 & 23,772 \\
J004054.7-091526    & 00 40 54.66 & -09 15 26.92  & 24.55 & 4.976  & 6700-8515,8655-10420 & 072.B-0123(D)Ê& 6 & 7,285  \\
QSO J0041-4936  & 00 41 31.44 & -49 36 11.80  & 17.90 &  3.240 & 3290-4520,4620-5600,5675-6650 & 68.A-0362(A) & 3 & 3,600\\
QSO B0039-407  & 00 42 01.20 & -40 30 39.00  & 18.50 & 2.478 & 3290-4520,4620-5600,5675-6650 & 075.A-0158(A)Ê& 15 & 16,500 \\
QSO B0039-3354    & 00 42 16.45 & -33 37 54.50  & 17.74 & 2.480 & 3295-4520 & 072.A-0442(A)Ê & 1 & 1,800 \\
LBQS 0041-2638     & 00 43 42.80 & -26 22 11.00  & 17.79 & 3.053 & 3757-4980,6700-8520,8660-10420 &70.A-0031(A)Ê & 6 & 5,400 \\
LBQS 0041-2707    & 00 43 51.81 & -26 51 28.00  & 17.83 & 2.786 & 3757-4980,6700-8520,8660-10420 & 70.A-0031(A)Ê & 6 & 5,400 \\
QSO B0042-2450    & 00 44 28.10 & -24 34 19.00 & 17.30 & 0.807 & 3300-4520,4620-5600,5675-6650 & 67.B-0373(A)Ê& 6 & 1,800  \\ 
QSO B0042-2656     & 00 44 52.26 & -26 40 09.12  & 19.00 & 3.358 & 3750-4980,6700-8520,8660-10420 & 70.A-0031(A)Ê& 6 & 3,600 \\
LBQS 0042-2930  & 00 45 08.54 & -29 14 32.59 & 17.92$^\dagger$ & 2.388 & 3290-4510,4780-5750,5830-6800 & 072.A-0442(A)Ê& 3 & 1,800  \\
LBQS 0042-2657    & 00 45 19.60 & -26 40 51.00 & 18.72 & 2.898 & 3757-4980,6700-8520,8660-10420 & 70.A-0031(A)Ê& 9 & 5,400  \\
J004612.2-293110  & 00 46 12.25 &  -29 31 10.17 & 19.89 & 1.675 & 3050-3870,4620-5600,5675-6650 & 077.B-0758(A)Ê& 26 & 45,470  \\
LBQS 0045-2606    & 00 48 12.59 & -25 50 04.80 & 18.08$^\dagger$ & 1.242 & 3300-4520,4615-5595,5670-6650 & 67.B-0373(A)Ê& 6 & 3,000  \\
QSO B0045-260    & 00 48 16.84 & -25 47 44.20 & 18.60 & 0.486 & 3300-4520,4615-5595,5670-6650 & 67.B-0373(A) & 6 & 7,200  \\
QSO B0046-2616    & 00 48 48.41 & -26 00 20.30 & 18.70 & 1.410 & 3300-4520,4615-5595,5670-6650 & 67.B-0373(A)Ê& 6 & 7,200  \\ 
LBQS 0047-2538  & 00 50 24.89 & -25 22 35.09 & 18.40 & 1.969 & 3290-4520,4615-5595,5670-6650 & 67.B-0373(A)Ê& 9 & 5,400  \\
LBQS 0048-2545    & 00 51 02.30 & -25 28 48.00 & 18.20 & 2.082 &  3295-4520,4615-5595,5670-6650 & 67.B-0373(A)Ê& 12 & 7,200  \\
QSO B0018-2608    & 00 51 09.10 & -25 52 15.00 & 18.20 & 2.249 & 3300-4520,4615-5595,5670-6650 & 67.B-0373(A)Ê& 6 & 3,600  \\
LBQS 0049-2535    & 00 52 11.10 & -25 18 59.00  & 19.20 & 1.528 & 3300-4520,4615-5595,5670-6650 & 67.B-0373(A)Ê& 12 & 7,200 \\
LBQS 0051-2605    & 00 54 19.78 & -25 49 01.20 & 18.04$^\dagger$ & 0.624 & 3300-4520,4615-5595,5670-6650 & 67.B-0373(A)Ê& 15 & 9,000  \\
QSO B0055-26  & 00 57 58.00 & -26 43 14.90 & 17.47 & 3.662 & 3050-5755,5835-6810 & 65.O-0296(A)Ê& 17 &  37,800  \\
QSO B0058-292  & 01 01 04.60 & -28 58 00.90 & 18.70 & 3.093 & 3300-5750, 5830-8515,8650-10420 & 67.A-0146(A), 66.A-0624(A)Ê& 29 & 41,222 \\
LBQS 0059-2735    & 01 02 17.05 & -27 19 49.91 & 18.00$^\dagger$ & 1.595 & 3100-5755,5830-8520,8660-10420 & 078.B-0433(A)Ê& 24 &  22,800  \\ 
QSO B0100+1300   & 01 03 11.27 & +13 16 17.74 & 16.57 & 2.686 & 3290-4525,6695-8520,8660-10420 & 074.A-0201(A)Ê, 67.A-0022(A)Ê& 7 & 12,600  \\
QSO J0105-1846  & 01 05 16.82 & -18 46 41.90 & 18.30 & 3.037 & 3300-5595,5670-6650,6690-8515, & 67.A-0146(A) & 9 & 14,400  \\
$\cdots$ &  &  & & & 8650-10420 & \\
QSO B0102-2931    & 01 05 17.95 & -29 15 11.41 & 19.90 & 2.220 & 3300-4515,5720-7520,7665-9460 & 075.A-0617(A) & 3 & 3,900  \\
QSO B0103 -260    & 01 06 04.30 & -25 46 52.30  & 18.31 & 3.365 & 4165-5160,5220-6210 & 66.A-0133(A)Ê& 16 & 24,800 \\
QSO B0109-353   & 01 11 43.62 & -35 03 00.40 & 16.90 & 2.406 & 3055-5755,5835-8515,8650-10420 & 166.A-0106(A) & 39 & 46,800   \\
QSO B0112-30  & 01 15 04.70 & -30 25 14.00 & $\cdots$ & 2.985 & 4778-5760,5835-6810 & 68.A-0600(A)Ê& 6 & 13,500 \\
QSO B0117+031    & 01 15 17.10 & -01 27 04.51 & 18.90 & 1.609 &  4165-5160,5230-6200 & 69.A-0613(A) & 2 & 3,600  \\
QSO J0123-0058    & 01 23 03.23 & -00 58 18.90 & 18.76 & 1.550 & 4165-5160,5230-6200 & 078.A-0646(A)Ê& 4 & 5,400   \\
QSO J0124+0044    & 01 24 03.78 & +00 44 32.67 & 17.90 & 3.834 & 4165-6810 & 69.A-0613(A) & 10 & 15,600   \\
QSO B0122-379    & 01 24 17.38 & -37 44 22.91 & 17.10 & 2.190 & 3050-5760,5835-8520,8660-10420 & 166.A-0106(A)Ê& 39 & 46,800  \\
QSO B0122-005  & 01 25 17.15 & -00 18 28.88 & 18.60 & 2.278 & 3300-4520,4620,5600,5675-6650 & 075.A-0158(A) & 12 & 13,200  \\
QSO B0128-2150  & 01 31 05.50 & -21 34 46.80 & 15.57 & 1.900 & 3045-3868,4785-5755,5830-6810 & 072.A-0446(B) & 6 & 6,130   \\
QSO B0130-403   & 01 33 01.93 & -40 06 28.00 & 17.40 & 3.023 & 3300-4520,4775-5755,5828-6801 & 70.B-0522(A)Ê& 3 & 3,065  \\
QSO J0133+0400    & 01 33 40.40 & +04 00 59.00 & 18.30 & 4.154 & 4180-5165,5230-6410, & 69.A-0613(A), 073.A-0071(A), & 5 & 9,100  \\
$\cdots$ &  &  & & & 6700-8520,8660-10250 & 074.A-0306(A)Ê \\
QSO J0134+0051    & 01 34 05.77 & 00 51 09.35 & 18.37 & 1.522 & 3050-3870,4620-5600,5675-6650 & 074.A-0597(A)Ê& 8 & 14,400  \\
QSO B0135-42  & 01 37 24.41 & -42 24 16.80 & 18.46 & 3.970 & 4170-5165,5235-6210 & 69.A-0613(A)Ê& 2 & 3,600  \\
QSO J0138-0005    & 01 38 25.54 & -00 05 34.52 & 18.97 & 1.340 & 3050-3870,4620-5598,5675-6650 & 078.A-0646(A)Ê& 9 & 15,600  \\
QSO J0139-0824    & 01 39 01.40& -08 24 44.08 & 18.65 & 3.016 & 3300-4520 & 074.A-0201(A)Ê& 1 &  4,800   \\
QSO J0143-3917    & 01 43 33.64 & -39 17 00.00 & 16.28 & 1.807 & 3050-5755,5830-8515,8655-10420 & 67.A-0280(A)Ê& 33 & 39,600  \\ 
QSO J0153+0009    & 01 53 18.19 & 00 09 11.39 & 17.80 & 0.837 & 3100-3870,4625-5600,5675-6650 & 078.A-0646(A)Ê& 6 & 10,800  \\
QSO J0153-4311 & 01 53 27.19 & -43 11 38.20 & 16.80 & 2.789 & 3100-5757,5835-8520,8650-10420 & 166.A-0106(A) & 53 & 63,900  \\
QSO J0157-0048    & 01 57 33.88 & -00 48 24.49 & 17.88 & 1.545 & 3300-4515,4785-5757,5835-6810 & 078.A-0003(A)Ê& 6 & 6,000  \\
QSO B0201+113  & 02 03 46.66 & +11 34 45.41 & 19.50 & 3.610 & 3760-4980,6700-8515,8660-10420 & 68.A-0461(A) & 3 & 3,000  \\
QSO J0209+0517  & 02 09 44.62 & +05 17 14.10 & 17.80  & 4.174 & 3300-4520,4775-5754,5830-6810 & 69.A-0613(A) & 3 & 3,600  \\
J021741.8-370100  & 02 17 41.78 & -37 00 59.60 & 18.40 & 2.910 & 3300-4520 & 072.A-0442(A)Ê& 2 & 7,200   \\
QSO J0217+0144   & 02 17 48.95 & +01 44 49.70  & 18.33 & 1.715 & 4165-5160,5230-6210 & 078.A-0646(A)Ê& 4 &  5,400 \\
QSO B0216+0803 & 02 18 57.36 & +08 17 27.43 & 18.10 & 2.996 & 3065-4520,4775-5755,5835-8525, & 073.B-0787(A), 072.A-0346(A)Ê & 30 & 37,200  \\
$\cdots$ & & & & & 8665-10250 & \\
QSO B0227-369 &  02 29 28.47 & -36 43 56.78 & 19.00 & 2.115 & 3300-4520,4620-5600,5675-6650 & 076.A-0389(A)Ê& 12 & 12,800   \\ 
QSO B0237-2322 & 02 40 08.18 & -23 09 15.78 & 16.78$^\dagger$  & 2.225 & 3050-5757,5835-8525,8660-10420 & 166.A-0106(A) & 78 & 91,149 \\
QSO J0242+0049   & 02 42 21.88 & +00 49 12.67 & 18.44 & 2.071 & 3300-4985,5715-7520,7665-9460 & 075.B-0190(A)Ê& 12 & 20,700  \\
\hline
\end{tabular}
\end{table*}

\addtocounter{table}{-1}
\begin{table*}
\caption{continued.}
\label{dla_log}
\centering
\setlength{\tabcolsep}{1pt}
\begin{tabular}{@{} l c c c c c c c c @{}}
\hline
 Quasar  & RA &  Dec & Mag. & $z_{\rm em}$ &  Wavelength coverage &  Prog. ID &  No. &  T$_{\rm exp}$ \\
  & 2000 & 2000 &  & & \AA\ &  & spec & sec  \\
\hline
QSO J0243-0550   & 02 43 12.47 & -05 50 55.30 & 19.00 & 1.805 & 3300-4518,4620-5600,5675-6650 & 076.A-0389(A)Ê& 9 & 6,180  \\ 
QSO B0241-01  & 02 44 01.84 & -01 34 03.70 & 17.06$^\star$  & 4.053 & 4165-6810 &  69.A-0613(A), 074.A-0306(A) & 4Ê& 5400 \\
QSO B0244-1249 & 02 46 58.47 & -12 36 30.80 & 18.40 & 2.201 & 3290-4520,4625-5600,5675-6650 & 076.A-0389(A)Ê& 6 & 6,000  \\
QSO B0253+0058 & 02 56 07.26 & +01 10 38.62 & 18.88 & 1.346 & 3300-4515,4620-5600,5675-6650 & 074.A-0597(A)Ê& 9 & 16,200  \\
QSO B0254-404   & 02 56 34.03 & -40 13 00.30 & 17.40 & 2.280 & 3292-4520 & 072.A-0442(A)Ê& 1 & 1,800  \\
QSO J0300+0048   & 03 00 00.57 & +00 48 28.00 & 19.48 & 0.314 & 3100-5755 & 267.B-5698(A)Ê& 9 & 16,492  \\
J030640.8-301032   & 03 06 40.93 & -30 10 31.91  & 19.32$^\dagger$ & 2.096 & 3050-5600,5670-6650,6700-8525, & 70.A-0031(A)Ê& 12 & 8,558 \\
$\cdots$ & & &&  &  8665-10420 \\
J030643.7-301107   & 03 06 43.77 & -30 11 07.58 & 19.85$^\dagger$ & 2.130 & 3060-5595,5675-6650,6700-8520, & 70.A-0031(A) & 15 & 12,600  \\
$\cdots$ & & &  & & 8665-10420 \\
QSO B0307-195A & 03 10 06.00 & -19 21 25.00 & 18.60 & 2.144 & 3050-5758,5835-8525,8660-10420 & 69.A-0204(A), 68.A-0216(A),Ê & 18 & 28,450  \\
$\cdots$ & & & && &  65.O-0299(A) & \\
QSO B0307-195B & 03 10 09.00 & -19 22 08.00 & 19.10 &  2.122 & 3065-5758,5835-8520,8660-10420 & 69.A-0204(A), 68.A-0216(A), & 27 & 37,700 \\
$\cdots$ & & &  & &  65.O-0299(A) & \\
J031856.6-060038   & 03 18 56.63 & -06 00 37.75  & 19.31 & 1.927 & 3100-5760,5835-8525,8665-10250 & 078.B-0433(A)Ê& 24 & 22,800 \\
QSO B0329-385   & 03 31 06.41 & -38 24 04.60 & 17.20 & 2.423 & 3060-5758,5835-8520,8655-10420 & 166.A-0106(A) & 36 & 31,200  \\
QSO B0329-2534 & 03 31 08.92 & -25 24 43.27 & 17.10  & 2.736 & 3100-5763,5830-8525,8665-10420 & 166.A-0106(A) & 66 & 77,363  \\
QSO J0332-4455 & 03 32 44.10 & -44 55 57.40 & 17.90 & 2.679 & 3300-4515,4780-5760,5840-6810 & 078.A-0164(A)Ê& 12 & 12,000  \\
QSO B0335-122   & 03 37 55.43 & -12 04 04.67  & 20.11 & 3.442 & 3755-4980,6700-8520,8660-10420 & 71.A-0067(A)Ê& 11 &  16,200 \\
QSO J0338+0021 & 03 38 29.31 &  +00 21 56.30 & 18.79$^\star$ & 5.020 & 6700-8525,8660-10420 & 072.B-0123(D)Ê& 6 & 10,450  \\
QSO J0338-0005   & 03 38 54.78 & -00 05 20.99 & 18.90 & 3.049 & 3100-3870 & 074.A-0201(A)Ê& 1 & 4,600  \\
QSO B0336-017   & 03 39 00.90 & -01 33 18.00 & 19.10 & 3.197 & 3757-4983,6700-8520,8660-10420 &  68.A-0461(A) & 9 & 15,600  \\ 
QSO B0347-383   & 03 49 43.68 & -38 10 31.10 & 17.30 & 3.222 & 3300-5757,5835-8520,8660-10420 & 68.B-0115(A)Ê& 18 & 28,800  \\
QSO B0347-2111   & 03 49 57.83 &  -21 02 47.74 & 21.10 & 2.944 & 3300-4520,4775-5755,5832-6810 & 71.A-0067(A)Ê& 9 & 8,873  \\
J035320.2-231418   & 03 53 20.10 & -23 14 17.80 & 17.00 & 1.911 & 3300-4515,4620-5600,5675-6650 & 078.A-0068(A) & 3 & 1,120  \\
QSO J0354-2724   & 03 54 05.56 & -27 24 21.00 & 17.90 & 2.823 & 3300-4515,4780-5760,5835-6810 & 078.A-0003(A)Ê& 12 & 10,240   \\
J040114.0-395132   & 04 01 14.06 & -39 51 32.70 & 17.00 & 1.507 & 3300-4515,4620-5600,5675-6650 & 078.A-0068(A) & 3 & 2,160  \\
QSO J0403-1703 & 04 03 56.60 & -17 03 23.00 & 18.70  & 4.227 & 4780-5757,5832-8525,8660-10420 & 074.A-0306(A) & 10 & 12,625  \\
QSO J0407-4410 & 04 07 18.08 & -44 10 14.10 & 17.60 &  3.020 & 3300-8520,8665-10420 & 68.A-0600(A), 70.A-0017(A),Ê& 54 & 94,000  \\
$\cdots$ & & & & & & 68.A-0361(A) & \\
QSO J0422-3844 & 04 22 14.79 & -38 44 52.90 & 16.90 & 3.123 & 3300-5600,5675-6650,6700-8515, & 166.A-0106(A) & 49 & 57,750  \\
$\cdots $ & & & & &8660-10420 &  \\
QSO J0427-1302   &  04 27 07.30 & -13 02 53.64 & 17.50 & 2.166 & 3285-4515,4780-5760,5835-6810 & 078.A-0003(A)Ê& 6 & 6,000  \\
QSO J0430-4855   & 04 30 37.31 & -48 55 23.70 & 16.20$^\dagger$ & 1.940 & 3060-5765,5845-8530,8665-10420 & 66.A-0221(A)Ê& 27 & 29,559  \\
QSO B0432-440   & 04 34 03.23 & -43 55 47.80 & 19.60 & 2.649 & 3300-4520,4780-5755,5830-6810 & 71.A-0067(A)Ê& 12 & 14,300  \\
QSO B0438-43   &  04 40 17.17 & -43 33 08.62  & 19.50  & 2.852 & 3100-5757,5835-8515,8660,10420 & 072.A-0346(A)Ê, 69.A-0051(A)Ê& 17 & 22,425 \\
QSO J0441-4313   & 04 41 17.32 & -43 13 45.40 & 16.40 & 0.593 & 3757-5758,5835-8515,8665-10420 &  70.C-0007(A) & 18 & 20,742   \\
QSO B0449-1645   & 04 52 13.60 & -16 40 12.00 & 17.00 & 2.679 & 3285-4515,4620-5600,5675-6650 & 078.A-0646(A)Ê& 9 & 15,600  \\
QSO B0450-1310B   & 04 53 12.80 & -13 05 46.00 & 16.50 & 2.250 & 3060-3874,4780-5760,5835-8525, &  70.B-0258(A)Ê& 15 & 15,000  \\
$\cdots$ & & & && 8670-10420 \\ 
QSO J0455-4216 & 04 55 23.05	& -42 16 17.40 & 17.30 & 2.661 & 3055-5758,5835-8520,8660-10420 & 166.A-0106(A) & 51 & 61,994  \\
PKS 0454-220  & 04 56 08.93 & -21 59 09.54  & 16.10 & 0.534 & 3050-3870,4170-5162,5230-6210 & 076.A-0463(A)Ê& 9 & 8,500 \\
4C-02.19 & 05 01 12.81 & -01 59 14.26  & 18.40 & 2.286 & 3100-8520,8660-10250 &074.B-0358(A), 66.A-0624(A), Ê& 26 & 42,422  \\
$\cdots$ & & & & && 076.A-0463(A) & \\
QSO B0512-3329 & 05 14 10.91 & -33 26 22.40 & 17.00 & 1.569 & 3300-4520,4620-5605,5675-6655 & 70.A-0446(A), 66.A-0087(A)Ê& 30 & 30,000  \\
QSO B0515-4414   & 05 17 07.63 & -44 10 55.50 & 14.90 & 1.713 & 3060-3875,4785-5760,5840-8525, & 66.A-0212(A)Ê & 9 & 13,500  \\
$\cdots$ & & & &&  8665-10420 & \\
J051939.8-364613   & 05 19 39.82 & -36 46 11.60 & 17.30 & 1.394 & 3290-4515,4620-5600,5675-6650 & 078.A-0068(A)Ê& 3 & 1,520  \\
QSO B0528-2505   & 05 30 07.96 & -25 03 29.84 & 17.30 & 2.765 & 3050-6810 & 68.A-0106(A), 66.A-0594(A), & 43 & 78,988  \\
$\cdots$ & & & & & & 68.A-0600(A)Ê& \\
QSO J0530+13  & 05 30 56.42 & +13 31 55.15 & 20.00 & 2.070 & 3300-5160,5230-6210,6705-8529, & 70.C-0239(A) & 18 & 12,600  \\
$\cdots$ & & & & & 8665-10420 & \\
QSO B0551-36   & 05 52 46.19	 & -36 37 27.60 & 17.00 & 2.318 & 3060-5757,5835-6810 & 66.A-0624(A) & 17 & 26,100  \\
J060008.1-504036  & 06 00 08.10 & -50 40 36.80  & 18.20$^\dagger$ & 3.130 & 3300-4520,4620-5600,5675-6650 & 68.A-0639(A)Ê& 20 & 24,000 \\
QSO B0606-2219   & 06 08 59.69 & -22 20 20.96 & 20.00 & 1.926 & 3300-4515,4625-5600,5675-6650 & 076.A-0389(A)Ê& 13 & 16,000  \\
QSO B0642-5038   & 06 43 27.00 & -50 41 12.80 & 18.50 & 3.090 & 3300-4520,4785-5757,5835-6810 & 073.A-0071(A)Ê& 9 & 17,500   \\
QSO B0736+01 & 07 39 18.03 &  +01 37 04.62 & 16.47 & 0.191 & 3300-5160,5230-6208,6715-8525, & 70.C-0239(A) & 26 & 17,600  \\
$\cdots$ & & & & & 8665-10420 & \\
QSO B0810+2554   & 08 13 31.29 & +25 45 03.06 & 15.40$^\dagger$ & 1.510 & 3050-3868,4625-5598,5675-6650 & 68.A-0107(A)Ê& 12 & 14,100  \\
QSO B0827+2421  & 08 30 52.08 & +24 10 59.82 & 17.30 & 0.939 & 3060-3872,4620-5600,5675-6650 & 68.A-0170(A), 69.A-0371(A)Ê& 23 & 34,071  \\
QSO B0841+129   & 08 44 24.20 & +12 45 48.90 & 18.50 & 2.505 & 3290-4520,6710-8525,8665-10420 & 70.B-0258(A)Ê& 9 &  10,800   \\
QSO B0908+0603 & 09 11 27.61 & +05 50 54.28  & 18.30$^\dagger$ & 2.793 & 3300-4522,4620-5600,5675-6650 & 70.A-0439(A)Ê& 39 & 46,800  \\
QSO B0913+0715 & 09 16 13.94 & +07 02 24.30 & 17.10 & 2.785 & 3280-10420 & 078.A-0185(A)Ê, 68.B-0115(A)Ê& 62 & 77,523 \\
QSO B0919-260   & 09 21 29.36 & -26 18 43.28 & 18.40 & 2.299 & 3300-4518,4618-5600,5675-6650  & 075.A-0158(A)Ê& 12 & 13,200  \\
QSO B0926-0201   & 09 29 13.57 & -02 14 46.40 & 16.44 & 1.661 & 3050-5757,5835-8525,8660-10420 & 072.A-0446(A)Ê& 15 & 15,325  \\
QSO B0933-333   & 09 35 09.23 & -33 32 37.69 & 20.00 & 2.910 & 3760-4984,6705-8525,8660-10420 & 71.A-0067(A), 69.A-0051(A)Ê& 12 & 14,300  \\
QSO B0951-0450   & 09 53 55.70 & -05 04 18.00 & 19.00 & 4.369 & 4785-5750,5832-6807 & 072.A-0558(A)Ê& 8 & 30,960  \\
QSO B0952+179 & 09 54 56.82 & +17 43 31.22 & 17.23  & 1.472 & 3060-3870,4618-5600,5675-6650 & 69.A-0371(A)Ê& 9 & 17,100 \\
QSO B0952-0115 & 09 55 00.10 & -01 30 07.00  & 18.70 & 4.426 & 3760-5754,5832-8525,8662-10420 & 70.A-0160(A), 072.A-0558(A) & 55 & 97,676 \\
QSO B1005-333 & 10 07 31.39 & -33 33 06.72 & 18.00 & 1.837 & 3300-4518,4620-5600,5675-6650 & 075.A-0158(B)Ê& 6 & 6,000  \\
QSO J1009-0026 & 10 09 30.47 & -00 26 19.13 & 17.53 & 1.241 & 3300-4515,4620-5600,5675-6650 & 078.A-0003(A)Ê& 6 & 6,000  \\
LBQS 1026-0045B   & 10 28 37.02 & -01 00 27.56 & 18.40 & 1.530 & 3300-4515,4620-5600,5675-6650 & 078.A-0003(A)Ê& 12 & 10,240  \\
\hline
\end{tabular}
\end{table*}

\addtocounter{table}{-1}
\begin{table*}
\caption{continued.}
\centering
\setlength{\tabcolsep}{1pt}
\begin{tabular}{@{} l c c c c c c c c@{}}
\hline
 Quasar & RA &  Dec &  Mag. & $z_{\rm em}$ &  Wavelength coverage &  Prog. ID &  No. &  T$_{\rm exp}$\\
 & 2000 & 2000 &  & & \AA\ &  & spec & sec  \\
\hline
QSO B1027+0540   &  10 30 27.10 & +05 24 55.00 & 18.87$^\star$  & 6.311 & 8000-10040 & 69.A-0529(A)Ê& 6 & 33,400  \\
Q1036-272 & 10 38 49.00 & -27 29 19.10 & 21.50 & 3.090 & 3757-4980,6700-8520,8660-10420 &70.A-0017(A)Ê & 9 & 12,133  \\
QSO B1036-2257 & 10 39 09.52 & -23 13 26.20 & 18.00 & 3.130 & 3300-5758,5838-8525,8660-10420 & 68.B-0115(A)Ê & 12 & 21,600  \\
QSO J1039-2719   & 10 39 21.86 & -27 19 16.45 & 17.40 & 2.193 & 3290-5600,5675-6650,6705-8525, & 69.B-0108(A), 70.A-0017(A)Ê& 30 & 36,500  \\
$\cdots$ & & & & & 8665-10420 & \\
QSO B1038-2712   & 10 40 32.23 & -27 27 49.00 & 17.70 & 2.331 & 3045-3865,4780-5757,5835-6810 & 67.B-0398(A), 65.O-0063(B)Ê& 9 & 16,200  \\ 
QSO B1036-268   & 10 40 40.32 & -27 24 36.40 & 20.00 & 2.460 & 3285-5600,5675-6650,6700-8520, & 67.B-0398(A) & 15 & 27,000  \\
$\cdots$ & & & & & 8660-10420 & \\
QSO J1044-0125   & 10 44 33.04 & -01 25 02.20 & 18.31$^\star$ & 5.740 & 8660-10420 & 268.A-5767(A), 69.A-0529(A)Ê& 7 & 30,270   \\
J104540.7-101813   & 10 45 40.56 & -10 18 12.80 & 17.40 & 1.261 & 3300-4515,4620-5600,5675-6650 & 078.A-0068(A)Ê& 3 & 1,650  \\
J104642.9+053107   & 10 46 42.84 & +05 31 07.02 & 18.00 & 2.682 & 3060-3870,4780-5755,5835-6810 & 074.A-0780(A) & 9 & 12,528  \\
QSO B1044+059   & 10 46 56.71 & +05 41 50.32 & 18.30 & 1.226 & 3060-3870,4780-5757,5835-6810 & 074.A-0780(A)Ê& 12 & 13,488  \\
QSO B1044+056   &  10 47 33.16 & +05 24 54.88 & 17.99 & 1.306 & 3060-3870,4780-5757,5835-6810 & 074.A-0780(A)Ê& 10 & 13,952  \\
QSO B1045+056   & 10 48 00.40 & +05 22 09.76 & 20.30 & 1.230 & 3060-3870,4780-5757,5830-6810 & 074.A-0780(A)Ê& 12 & 20,932  \\
QSO B1052-0004   & 10 54 40.98 & -00 20 48.47  & 18.51 & 1.021 & 3300-4515,4620-5600,5675-6650 & 078.A-0003(A)Ê& 12 & 10,240 \\
QSO B1055-301 & 10 58 00.43 & -30 24 55.03 & 19.50 & 2.523 & 3300-4520,4780-5757,5840-6810 & 71.A-0067(A) & 9 & 10,725  \\
QSO B1101-26   & 11 03 25.31 & -26 45 15.88 & 16.00 &  2.145 & 3780-4985,6710-8525,8665-10250 & 076.A-0463(A)Ê & 27 & 60,672  \\
QSO B1104-181 &  11 06 33.39 & -18 21 23.80 & 15.90 & 2.319 & 3060-8520,8655-10420 & 67.A-0278(A)Ê& 51 & 60,948  \\
QSO J1107+0048   & 11 07 29.03 & +00 48 11.27 & 17.66  & 1.392 & 3300-4515,4620-5600,5675-6650 & 074.A-0597(A)Ê& 6 & 7,200  \\
QSO B1108-07   & 11 11 13.60 & -08 04 02.00 & 18.10 & 3.922 & 3760-8520,8660-10420 & 67.A-0022(A), 68.B-0115(A), & 14 & 28,498  \\
 $\cdots$ & & & & & & 68.A-0492(A)Ê& \\
QSO J1113-1533   & 11 13 50.61 & -15 33 33.90 & 18.70 & 3.370 & 3760-5600,5675-6650 & 68.A-0492(A)Ê& 10 & 25,200  \\
QSO B1114-220   & 11 16 54.50 & -22 16 52.00 & 20.20  & 2.282 & 3300-4520,4620-5600,5675-6650 & 71.B-0081(A)Ê& 30 & 14,115 \\
QSO B1114-0822   & 11 17 27.10 & -08 38 58.00 & 19.40 & 4.495 & 4780-5757,5835-8520,8665-10250 & 074.A-0801(B) & 9 & 28,525  \\
QSO B1122-168 & 11 24 42.87 & -17 05 17.50 & 16.50 & 2.400 &  3290-4520,4620-5600,5675-6650 & 68.A-0570(A), 073.B-0420(A)Ê& 82 & 104,493   \\
J112910.9-231628   & 11 29 10.86 & -23 16 28.17 & 17.30 & 1.019 & 3300-4520,4620-5600,5675-6650 & 078.A-0068(A)Ê& 3 & 1,470 \\
QSO J1142+2654   & 11 42 54.26 & +26 54 57.50 & 17.00 & 2.630 & 3757-4980,6705-8520,8660-10420 & 69.A-0246(A) & 42 & 48,696  \\
QSO B1145-676 & 11 47 32.40 & -67 53 42.70 & 18.50$^\dagger$ & 0.210 & 3300-5160,5230-6210,6700-8525, & 70.C-0239(A) & 16 & 10,800  \\
$\cdots$ & & & & &  8660-10420 & \\
QSO B1151+068 & 11 54 11.12 & +06 34 37.80 & 18.60 & 2.762 & 3060-5757,5835-8520,8660-10420 & 65.O-0158(A) & 18 & 21,600  \\
J115538.6+053050    & 11 55 38.60 &  +05 30 50.67 & 19.08 & 3.475 & 3300-5600,5675-7500,7665-9460 & 076.A-0376(A)Ê& 11 & 14,400  \\
QSO J1159+1337   & 11 59 06.52 & +13 37 37.70 & 18.50 & 3.984 & 4780-5757,5835-6810 & 074.A-0306(B)Ê& 2 & 3,000  \\
QSO B1158-1842 & 12 00 44.94 & -18 59 44.50 & 16.90 & 2.453 & 3050-5757,5835-8520,8660-10420 & 166.A-0106(A)Ê& 36 & 43,200  \\
QSO B1202-074  & 12 05 23.12 & -07 42 32.40 & 17.50 & 4.695 & 3300-4520,4785-5765,5835-8525, &  71.B-0106(A), 66.A-0594(A),  & 43 & 82,086   \\
 $\cdots$ & & & & & 8665-10420 &  166.A-0106(A) \\
J120550.2+020131 & 12 05 50.19 & +02 01 31.55 & 17.46 &  2.134 & 3290-4520,4780-5757,5840-6810 & 273.A-5020(A)Ê& 3 & 3,000  \\
QSO B1209+0919 & 12 11 34.95 & +09 02 20.94 & 18.50 & 3.292 & 3300-8515,8650,10420 &  073.B-0787(A), 67.A-0146(A)Ê& 15 & 23,765   \\
LBQS 1209+1046  & 12 11 40.59 & +10 30 02.03 & 17.80 & 2.193 & 3290-4520,4620-5600,5675-6650 & 68.A-0170(A)Ê& 12 & 14,400  \\
LBQS 1210+1731  & 12 13 03.03 & +17 14 23.20 & 17.40 & 2.543 & 3060-3875,6705-8525,8665-10420 & 70.B-0258(A) & 21 & 30,500  \\
QSO J1215+3309   & 12 15 09.22 & +33 09 55.23 & 16.50 & 0.616 & 3300-4520,4620-5600,5675-6650 & 69.A-0371(A)Ê& 12 & 18,468  \\
QSO B1220-1800 & 12 23 10.62 & -18 16 42.40 & 18.10 & 2.160 & 3280-4520,4620-5600,5675-6650 & 273.A-5020(A)Ê& 3 & 3,000  \\
LBQS 1223+1753   & 12 26 07.20 & +17 36 49.90 & 18.10 & 2.940 & 3300-5600,5675-6650,6700-8520, & 69.B-0108(A), 65.O-0158(A)Ê& 14 & 18,000  \\
 $\cdots$ & & & & & 8660-10420 & \\
QSO B1228-113   & 12 30 55.54 & -11 39 09.62 & 22.01$^\dagger$ & 3.528 & 3300-4520,4780-5760,5835-6810 & 71.A-0067(A)Ê& 14 & 17,875  \\
QSO B1230-101   & 12 33 13.16 & -10 25 18.44 & 19.80 & 2.394 & 3300-4515,4620-5600,5675-6650 & 075.A-0158(A)Ê& 12 & 12,000  \\
LBQS 1232+0815   & 12 34 37.55 & +07 58 40.50 & 18.40 & 2.570 & 3285-4520,4620-5600,5675-6650 & 68.A-0106(A), 69.A-0061(A), & 36 & 59,400  \\
LBQS 1242+0006   &  12 45 24.60 & -00 09 38.01 & 17.70 & 2.084 &  3290-4520,4620-5600,5675-6650 & 273.A-5020(A)Ê& 3 & 3,000  \\
QSO J1246-0730   & 12 46 04.24 & -07 30 46.61 & 18.00 & 1.286 & 3050-3870,4780-5758,5835-6810 & 69.A-0410(A)Ê& 12 & 12,000  \\
LBQS 1246-0217 & 12 49 24.87 & -02 33 39.76 & 18.10 & 2.117 & 3285-4520,4620-7500,7665-9460 & 076.A-0376(A), 67.A-0146(A)Ê & 21 & 27,000  \\
J124957.2-015929   & 12 49 57.24 & -01 59 28.76 & 18.58 & 3.635 & 3300-4515,4775-5757,5835-6810 & 075.A-0464(A) & 33 & 33,650   \\
QSO B1249-02 & 12 51 51.39 & -02 23 33.60 & 17.10 & 1.192 & 3300-4520,4620-5600,5675-6650 & 078.A-0068(A)Ê& 3 & 1,218  \\
QSO B1256-177  & 12 58 38.29 & -18 00 03.18 & 20.20 & 1.956 & 3290-4520,4620-5600,5675-6650 & 075.A-0158(A) & 12 & 12,000  \\
QSO J1306+0356   & 13 06 08.26 & +03 56 26.30 & 18.77$^\star$ & 5.999 & 6700-10420 & 69.A-0529(A), 077.A-0713(A)Ê& 8 & 38,131  \\
QSO B1317-0507   & 13 20 29.98 & -05 23 35.50 & 16.54 & 3.710 & 3300-4520,4775-5757,5835-6810 & 075.A-0464(A)Ê& 23 &  24,040  \\
QSO B1318-263 & 13 21 14.06 & -26 36 11.19 & 20.40$^\dagger$ & 2.027 & 3300-4520,4620-5600,5675-6650 & 075.A-0158(A)Ê& 20 & 23,100  \\
LBQS 1320-0006 & 13 23 23.79 & -00 21 55.24 &18.20  & 1.388 & 3300-4520,4620-5600,5675-6650 & 274.A-5030(A)Ê& 15 & 13,841  \\
QSO B1324-047   & 13 26 54.61 & -05 00 58.98 & 19.00 & 1.882 & 3290-4520,4620-5600,5675-6650 & 075.A-0158(A)Ê& 17 & 19,800  \\
QSO J1330-2522   & 13 30 51.98 & -25 22 18.80 & 18.46 & 3.910 & 3300-4515,4780-5757,5835-6810 & 077.A-0166(A)Ê& 3 & 5,400  \\
QSO B1331+170 & 13 33 35.78 & +16 49 03.94 & 16.71 & 2.084 & 3060-4520,4620-5600,5675-5660, & 67.A-0022(A), 68.A-0170(A)Ê& 15 & 20,700  \\
 $\cdots$ & & & & & 6692-8515,8650-10420 & \\
QSO J1342-1355   & 13 42 58.86 & -13 55 59.78 & 19.20 & 3.212 & 3300-5600,5675-6650 & 68.A-0492(A)Ê& 10 & 25,100  \\
QSO J1344-1035   & 13 44 27.07 & -10 35 41.90 & 17.10 & 2.134 & 3050-5760,5835-8520,8660-10420 & 166.A-0106(A)Ê& 51 & 58,933   \\
QSO B1347-2457 & 13 50 38.88 & -25 12 16.80 & 16.30 & 2.578 & 3050-5765,5835-8525,8660-10420 & 166.A-0106(A)Ê& 30 & 35,444  \\ 
QSO J1356-1101   & 13 56 46.83 & -11 01 29.22 & 19.20 & 3.006 & 3757-4985,6700-8520,8660-10420 & 71.A-0067(A), 71.A-0539(A), & 18 & 13,725  \\
 $\cdots$ & & & & & &  69.A-0051(A) & \\
QSO B1402-012 & 14 04 45.89 & -01 30 21.84 & 18.38 & 2.522 & 3757-4985,5710-7518,7665-9460 & 075.A-0158(B)Ê& 6 & 5,300  \\
 $\cdots$ & & & & & & 71.B-0136(A) \\
QSO B1409+0930 & 14 12 17.30 &  +09 16 25.00 & 18.60 & 2.838 & 3050-5757,5830-8520,8665-10420 & 65.O-0158(A), 69.A-0051(A)Ê& 44 & 51,950  \\
QSO B1412-096   & 14 15 20.83 & -09 55 58.33 & 17.50 & 2.001 & 3300-4515,4620-5600,5675-6650 & 075.A-0158(B)Ê& 3 & 3,300  \\
QSO J1421-0643 & 14 21 07.76 & -06 43 56.30  & 18.50 & 3.689 & 3757-4980,6705-8520,8665-10420 & 71.A-0067(A), 71.A-0539(A) & 20 & 18,120 \\
QSO B1424-41 & 14 27 56.30 & -42 06 19.55 & 17.70 & 1.522 & 3300-5160,5235-6210,6700-8520, & 67.C-0157(A) & 10 & 8,000  \\
 $\cdots$ & & & & & 8660-10420 & & \\
QSO B1429-008B &14 32 28.95 & -01 06 13.55 & 20.00$^\star$ & 2.082 & 3300-5600,5675-6650-6700-8525, & 69.A-0555(A)Ê& 49 &  58,048  \\
\hline
\end{tabular}
\end{table*}

\addtocounter{table}{-1}
\begin{table*}
\caption{continued.}
\centering
\setlength{\tabcolsep}{1pt}
\begin{tabular}{@{} l c c c c c c c c@{}}
\hline
 Quasar & RA &  Dec & Mag. &  $z_{\rm em}$ &  Wavelength coverage &  Prog. ID &  No. &  T$_{\rm exp}$ \\
 & 2000 & 2000 &  & & \AA\ &  & spec & sec  \\
\hline
 $\cdots$ & & & & & 8660-10420 & & \\
QSO J1439+1117   & 14 39 12.04 & +11 17 40.49 & 16.56$^\star$ & 2.583 & 3300-4515,4780-7105 & 278.A-5062(A)Ê& 23 & 34,273  \\
QSO J1443+2724   & 14 43 31.17 & +27 24 36.77 & 19.30 & 4.443 & 4780-5757,5835-7915,9900 & 077.A-0148(A), 072.A-0346(B)Ê& 28 &71,402  \\
LBQS 1444+0126 & 14 46 53.04 & +01 13 56.00 & 18.50 & 2.210 & 3290-4530,4620-5600-5675-6650 & 67.A-0078(A), 69.B-0108(A), Ê& 36 & 54,000  \\
 $\cdots$ & & & & & & 65.O-0158(A)Ê, 71.B-0136(A) & \\
QSO B1448-232 & 14 51 02.51 & -23 29 31.08 & 16.96 & 2.215 & 3050-8515,8660-10420 & 077.A-0646(A), 166.A-0106(A)Ê& 60 & 66,532  \\
J145147.1-151220    & 14 51 47.05 & -15 12 20.10 & 19.14 &  4.763 & 3755-5757,5835-8515,8655,10420 & 166.A-0106(A)Ê& 31 & 48,400  \\
QSO J1453+0029   & 14 53 33.01 & +00 29 43.56 & 21.46 & 1.297 & 4780-5755,5830-8550,8650-10420 & 267.B-5698(A)Ê& 8 & 18,000  \\
J151352.52+085555   & 15 13 52.53 & +08 55 55.74 & 15.57$^\star$ & 2.904 & 3300-4520,4620-5600,5675-6650 & 69.B-0108(A), 71.B-0136(A)Ê & 15 & 15,514  \\
QSO J1621-0042   & 16 21 16.92 & -00 42 50.90 & 17.23 & 3.700 & 3300-4515,4780-5757,5835-6810 & 075.A-0464(A)Ê& 23 & 19,096  \\
4C 12.59 & 16 31 45.16 & +11 56 03.01 & 18.50 & 1.792 & 3060-3870,4780-5757,5835-6810 & 69.A-0410(A)Ê & 12 & 12,000  \\
QSO J1723+2243    & 17 23 23.10 & +22 43 56.90 & 18.17 & 4.520 & 4780-5757,5835-6810 & 073.A-0071(A)Ê& 2 & 4,500  \\
QSO B1730-130    & 17 33 02.72 & -13 04 49.49 & 18.50  & 0.902 & 3300-5160,5230-6210,6695-8515, & 67.C-0157(A) & 10 & 9,600 \\
 $\cdots$ & & & & & 8655-10420 & \\
QSO B1741-038    & 17 43 58.86 & -03 50 04.62 & 18.50 & 1.054 & 3300-5160,5230-6210,6690-8515, & 67.C-0157(A) & 13 &  12,000  \\
 $\cdots$ & & & & & 8650-10420 & \\
QSO B1937-1009    & 19 39 57.25 & -10 02 41.54 & 19.00 & 3.787 & 4780-5760,5835-6810 & 077.A-0166(A)Ê& 10 & 27,000   \\
QSO B1935-692    & 19 40 25.51 & -69 07 56.93 & 18.80 & 3.152 & 4780-5757,5835-6810 & 65.O-0693(A)Ê& 10 & 23,512 \\
QSO B2000-330   & 20 03 24.12 & -32 51 45.03 & 18.40 & 3.783 & 3760-8535,8650-10420  & 166.A-0106(A), 65.O-0299(A)Ê& 43 & 75,600   \\
QSO J2107-0620    & 21 07 57.67 & -06 20 10.66 & 17.49 & 0.642 & 3757-4985 & 075.B-0190(A)Ê& 3 & 5,400   \\
LBQS 2113-4345 & 21 16 54.25 & -43 32 34.00 & 18.50  & 2.053 & 3050-7500,7660-9460 & 69.A-0586(A), 077.A-0714(A) & 17 & 10,536  \\
LBQS 2114-4347    & 21 17 19.34 & -43 34 24.40 & 18.30 & 2.040 & 3050-10420 & 69.A-0586(A), 077.A-0714(A) & 29 & 22,325  \\
J211739.5-433538    & 21 17 39.46 & -43 35 38.50 & 18.30$^\dagger$ & 2.050 & 3050-5757,5835-8520,8660-10420 & 69.A-0586(A) & 14 & 10,300  \\
QSO J2119-3536 & 21 19 27.60 & -35 37 40.60 & 17.00 & 2.341 & 3290-4530,4620-5600,5675-6650 & 65.O-0158(A)Ê& 6 & 7,200  \\
QSO B2126-15   & 21 29 12.17 & -15 38 41.17 & 17.30 & 3.268 & 3300-5600,5675-6650,6695-8520, & 166.A-0106(A)Ê& 87 & 102,302  \\
 $\cdots$ & & & & & 8650-10420 & \\
QSO B2129-4653   & 21 33 02.10 & -46 40 28.80 & 18.90 & 2.230 & 3290-4515,4620-5600,5675-6650 &  074.A-0201(A)Ê & 6 & 7,200  \\
J213314.2-464031   & 21 33 14.17 & -46 40 31.80 & 19.96$^\dagger$ &  2.208 & 3290-4520,4620-5600,5675-6650 & 073.A-0071(A)Ê& 9 & 18,000  \\
LBQS 2132-4321 & 21 36 06.04 & -43 08 18.10 & 17.68 & 2.420 & 3290-4530,4620-5600,5675-6650 & 65.O-0158(A) & 3 &  3,600  \\
LBQS 2138-4427   & 21 41 59.79 & -44 13 25.90 & 18.20 & 3.170 & 3757-4983,6692-8515,8650-10420 & 67.A-0146(A)Ê & 9 & 15,120  \\
QSO B2139-4433   &  21 42 22.23 & -44 19 29.70 & 20.18 & 3.220 & 4778-5757,5835-6810 & 69.A-0204(A)Ê & 8 & 21,600  \\
QSO B2149-306   & 21 51 55.52 & -30 27 53.63 &18.00 & 2.345 & 3300-4518,4620-5600,5675-6650 & 075.A-0158(B)Ê& 6 & 5,200   \\
QSO B2204-408   & 22 07 34.41 & -40 36 56.00 & 17.50 & 3.155 & 4778-5755,5835-6810 & 71.B-0106(B)Ê& 6 & 9,900  \\
LBQS 2206-1958A & 22 08 52.07 & -19 44 00.00 & 17.33 &  2.560 & 3060-8520,8665-10420 & 65.O-0158(A), 072.A-0346(A)Ê& 27 & 33,900  \\
QSO J2215-0045   &  22 15 11.94 & -00 45 50.01 & 17.40 & 1.476 & 3060-5757,5830-8515,8650-10420 & 267.B-5698(A)Ê& 18 & 21,600  \\
QSO J2220-2803 & 22 20 06.76 & -28 03 23.34 & 16.00 & 2.406 & 3060-8520,8655-10420 & 166.A-0106(A), 65.P-0183(A)Ê& 60 & 72,000  \\
QSO B2222-396   & 22 25 40.44 & -39 24 36.66 & 17.90 & 2.198 & 3290-4520 &  072.A-0442(A)Ê& 1 & 1,800  \\
QSO J2227-2243   & 22 27 56.94 & -22 43 02.60 & 17.60 & 1.891 & 3060-5757,5835-8515,8660-10420 & 67.A-0280(A)Ê& 36 & 41,813   \\
QSO B2225-4025   & 22 28 26.95 & -40 09 58.90 & 18.10  & 2.030 & 3300-4520 & 072.A-0442(A)Ê& 1 & 1,800  \\
LBQS 2230+0232   & 22 32 35.23 & +02 47 55.80 & 18.04 & 2.147 & 3060-3872,6705-8525,8665-10420 & 70.B-0258(A)Ê& 9 & 13,500  \\
J223851.0-295301   & 22 38 50.97 & -29 53 00.59 & 19.53$^\dagger$ & 2.387 & 3080-5750,5830-8520,8660-10420 & 69.A-0586(A) & 14 & 12,180  \\
J223922.9-294947   & 22 39 22.86 & -29 49 47.73 & 19.59$^\dagger$ & 1.849 & 3060-5757,5835-6810 & 69.A-0586(A)Ê& 11 & 10,708 \\
J223938.9-295451   & 22 39 38.91 & -29 54 50.62 & 19.66$^\dagger$ & 1.907 & 3080-7505,7670-9460 & 69.A-0586(A), 077.A-0714(A) & 28 & 29,190  \\
J223941.8-294955    & 22 39 41.76 & -29 49 54.54 & 19.31$^\dagger$ & 2.102 & 3060-7500,7670-9460 & 69.A-0586(A), 077.A-0714(A) & 22 & 21,626  \\
J223948.7-294748   & 22 39 48.67 & -29 47 48.18 & 20.19$^\dagger$ & 2.068 & 3060-10420 & 69.A-0586(A), 077.A-0714(A) & 33 & 31,590  \\
J223951.9-294837   & 22 39 51.84 & -29 48 36.52 & 18.85$^\dagger$ & 2.121 & 3070-5757,5835-6810 & 077.A-0714(A) & 11 & 8,450  \\
QSO B2237-0607  & 22 39 53.66 & -05 52 19.90 & 18.30 & 4.558 & 4775-5753,5830-6810 & 69.A-0613(A)Ê& 2 & 3,600  \\
QSO J2247-1237   & 22 47 52.64 & -12 37 19.72 & 18.50 & 1.892 & 3300-4518,4620-5600,5675-6650 & 075.A-0158(B)Ê& 9 & 7,200  \\
QSO B2311-373    & 23 13 59.71 & -37 04 45.20 & 18.50 & 2.476 & 3290-4520,4780-5757,5835-6810 & 69.A-0051(A)Ê& 9 & 10,725  \\
J232046.7-294406   & 23 20 46.72 & -29 44 06.06 & 19.83$^\dagger$ & 2.401 & 3050-5757,5835-8520,8660-10420 & 69.A-0586(A)Ê& 12 & 6,050  \\
J232059.4-295520   & 23 20 59.41 & -29 55 21.49 & 18.99$^\dagger$ & 2.317 & 3060-5757,5835-8520,8660-10420 &  69.A-0586(A)Ê&  12 & 6,050  \\
J232114.3-294725   & 23 21 14.25 & -29 47 24.29 & 19.99$^\dagger$ & 2.677 & 3758-4982,6700-8525,8665-10420 & 70.A-0031(A)Ê& 6 & 4,800  \\
J232121.2-294350   & 23 21 21.31 & -29 43 51.16 & 19.70$^\dagger$ & 2.184 & 3060-5757,5835-8520,8660-10420 &  69.A-0586(A)Ê& 12 & 9,155  \\
QSO B2318-1107   & 23 21 28.80 & -10 51 21.20   & $\cdots$ & 2.960 & 3050-4515,4780-5760,5840-6810 & 072.A-0442(A), 073.A-0071(A) & 13 & 23,400  \\
QSO J2328+0022   & 23 28 20.38 & +00 22 38.26 &17.95 & 1.302 & 3300-4520,4620-5600,5675-6650 & 074.A-0597(A)Ê& 9 & 15,900  \\
QSO B2332-094   & 23 34 46.40 & -09 08 12.33 & 18.66 & 3.330 & 3757-5757,5835-8520,8660-10250 & 68.A-0600(A), 073.B-0787(A)Ê& 25 & 31,920  \\
J233544.2+150118   & 23 35 44.19 & +15 01 18.33  & 18.20 & 0.791 & 3060-3870,4620-5600,5675-6650 & 078.A-0646(A)Ê& 12 & 17,920 \\
QSO B2342+3417 & 23 44 51.26 & +34 33 48.74 & 19.10  & 3.010 & 3757-4985,6705-8525,8665-10420 & 71.A-0539(A)Ê& 24 & 12,000 \\
QSO J2346+1247   & 23 46 25.40 & +12 47 44.00 & $\cdots$ & 2.578 & 3300-4518,4620-5600,5675-6650 & 075.A-0018(A)Ê& 42 & 70,240  \\
QSO B2343+125 & 23 46 28.22 & +12 48 59.93 & 17.50 & 2.763 & 3060-3872,4780-5757,5835-8515, & 67.A-0022(A) & 19 & 25,200  \\
 $\cdots$ & & & & &  & 8650-10420 & &  \\
QSO B2345+000   & 23 48 25.37 & +00 20 40.11 & 19.3 & 2.654 & 3300-4518,4620-5600,5675-6650 &  076.A-0320(A) & 24 & 25,600 \\
QSO B2347-4342   & 23 50 34.27 & -43 25 59.70 & 16.30 & 2.885 & 3070-5757,5835-8520,8655-10420 & 71.A-0066(A), 166.A-0106(A),Ê & 71 & 125,443  \\
 $\cdots$ & & & & & &  68.A-0230(A) \\
QSO B2348-0180   & 23 50 57.89 & -00 52 10.01 & 19.50 & 3.023 & 3300-4520,4780-5755,5835-6810 & 072.A-0346(A)Ê& 9 & 16,200  \\
QSO B2348-147    & 23 51 29.81 & -14 27 56.90 & 16.90 & 2.933 & 3290-4520,6100-7928,8070-9760 & 70.B-0258(A) & 9 & 10,350  \\
J235534.6-395355   & 23 55 34.61 & -39 53 55.30 & 16.30 & 1.579 & 3300-4515,4620-5600,5675-6650 &  078.A-0068(A) & 3 & 612  \\
J235702.5-004824   & 23 57 02.55 & -00 48 24.00 & 19.53 & 3.013 & 3300-4986 & 075.B-0190(A) & 7 & 25,200  \\
QSO J2359-1241   & 23 59 53.64 & -12 41 48.20 & 16.7 & 0.868 & 3050-5757,5835-8525,8665-10250 & 078.B-0433(A)Ê& 24 & 22,800 \\ [3pt]
\hline
Total  & $\cdots$ & $\cdots$ & $\cdots$ & $\cdots$ & $\cdots$ & $\cdots$ & 4309 & 5,616,229 \\
\hline
\end{tabular}
\vspace{0.1cm}
$\dagger$ $B$-band magnitude\\
$\star$ $J$-band magnitude
\end{table*}

\end{document}